\documentclass[12pt]{spieman}  
\usepackage{amsmath,amsfonts,amssymb}
\usepackage{graphicx}
\usepackage{setspace}
\usepackage{tocloft}
\usepackage[skip=10pt]{caption}
\usepackage{float}

\usepackage{subcaption}

\usepackage{algorithm}
\usepackage{algpseudocode}

\title{Deformable Registration of MRA and 4D Flow Images to Facilitate Accurate Estimation of Flow Properties within Blood Vessels}

\author[a]{Dan Lior}
\author[a]{Craig G. Rusin}
\author[a]{Justin Weigand}
\author[c]{Kristina V. Montez}
\author[d]{Yimo Wang}
\author[a]{Silvana Molossi}
\author[a]{Daniel J. Penny}
\author[b]{Charles Puelz}

\affil[a]{Department of Pediatrics, Division of Cardiology, Baylor College of Medicine and Texas Children's Hospital, Houston, TX, USA}
\affil[b]{Department of Mathematics, University of Houston, Houston, TX, USA}
\affil[c]{School of Engineering Medicine, Texas A\&M University, Houston, TX, USA}
\affil[d]{Department of Computational Applied Mathematics and Operations Research, Rice University, Houston, TX, USA}

\cftpagenumbersoff{figure}
\cftpagenumbersoff{table} 

\begin{document} 
\maketitle

\begin{abstract}
A method is presented for the registration of MRA and 4D Flow images, with the goal of calculating blood flow properties using both modalities simultaneously. In particular, the method produces an alignment of segmentations of vessel networks, from MRA images, with the blood velocity field within those networks, from the corresponding 4D Flow images. The alignment procedure is driven by the registration of centerlines of vessels extracted from the two modalities. Our approach is robust to noise, small deformations, and partial omissions of vessel surfaces and/or blood velocities. The alignment procedure is tested on 7 patient data sets acquired at Texas Children's Hospital. The quality of the resulting alignment is assessed by $(i)$ an illustration of the aligned and unaligned surface segmentations for a sample patient, $(ii)$ histograms of distances between centerline networks, and $(iii)$ graphs of estimated blood flow. For each of the 7 analyzed data sets, medians of the distance histograms decreased an average of 83.5\%, and the estimated blood flow increased significantly as a result of the alignment procedure. 

\end{abstract}


{\noindent \footnotesize\textbf{*}Dan Lior, \linkable{dan.lior@bcm.edu} }, 
{\noindent \footnotesize\textbf{*}Charles Puelz, \linkable{cepuelz@central.uh.edu} }

\begin{spacing}{1} 

\section{Introduction} \label{sec:intro}



Understanding the relationship between vascular geometry and flow properties through major blood vessels has important clinical implications, especially in the treatment of congenital heart disease. For example, surgical reconstruction of the aorta in single ventricle physiology alters vessel morphology and confers disturbed flow patterns. In turn, altered flow may affect single ventricle function, hemodynamics, and clinical outcomes \cite{Fogel2014-ql,Biglino2012-lm,Biglino2014-df, BIKO2019574, Schafer2019-zj, Bruse2017-cl, Taylor-LaPole2023-jp}. These ideas suggest that the analysis of flow properties (e.g. flow rate, wall shear stress, pulse wave velocity, and energy loss) can be used to assess both disease severity and surgical success.

Patients with congenital heart disease often undergo magnetic resonance imaging to quantify the geometry and blood flow patterns within their major blood vessels. Magnetic resonance angiography (MRA) provides high-resolution measurements of water content, from which a segmentation of the vessel lumen can be extracted. Four-dimensional phase-contrast magnetic resonance (4D Flow) provides time-resolved fluid velocity measurements. Estimates of blood flow properties within a vessel rely on measurements of the blood velocity within the vessel, captured by the 4D Flow sequence, and on a geometric representation of the vessel lumen itself, captured by the MRA sequence. It should be noted that segmentations of the vessel lumen can be obtained from the 4D Flow image, but they tend to be inaccurate, incomplete, and biased. These properties likely cause challenges in processing 4D Flow images, as demonstrated by Oechtering using various post processing software \cite{Oechtering2023-rz}, and are the main motivation for the present work.

First, the {\em inaccuracy} of vessel segmentations extracted from 4D Flow images comes from the relatively coarse spatial resolution of such images. Second, since the 4D Flow signal is blind to fluid flow below a certain speed, some vessels are only visible during certain intervals of the cardiac cycle, and some vessels, exhibiting little to no flow, are entirely unrepresented in the 4D Flow image. This leads to {\em incompleteness} of the vessel segmentations. Finally, segmentations of the vessel lumen acquired from 4D Flow images correspond to the boundary between high and low fluid velocities. A common assumption in the modeling of fluid flow through a vessel, the \emph{no-slip condition}, states that the velocity of fluid at the vessel boundary is zero. Due to the no-slip condition, this boundary invariably occurs strictly inside the vessel wall, thus introducing a {\em bias} towards narrower vessels in the segmented surfaces.

\begin{figure}[!ht]
\centering
\begin{subfigure}{0.49\textwidth}
   \scalebox{-1}[1]{
       \includegraphics[width=1\linewidth]{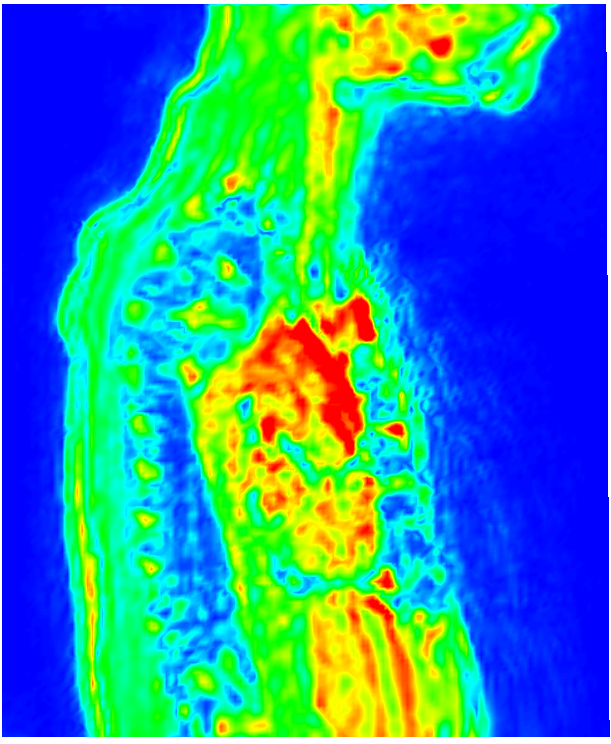} 
   }
  \caption{}
\end{subfigure}
\hspace{.5cm}
\begin{subfigure}{0.45\textwidth}
   \scalebox{-1}[1]{
       \includegraphics[width=1\linewidth]{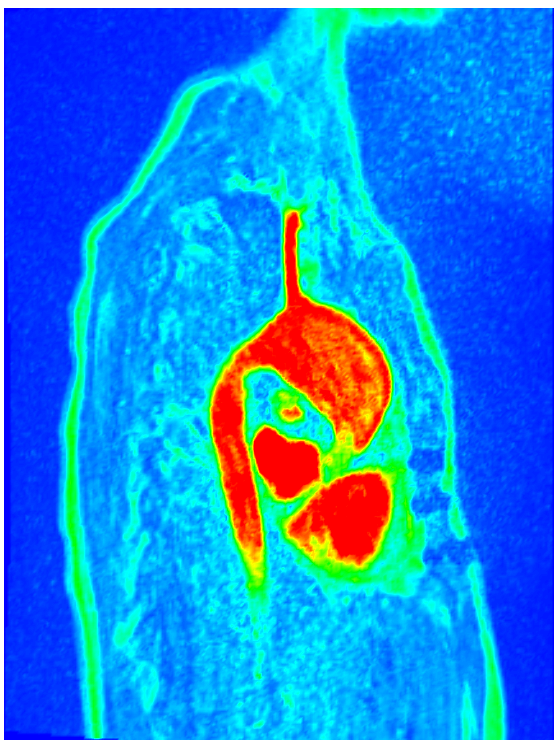} 
   }
  \caption{}
\end{subfigure}
\caption{Sections of the aorta as depicted in (a) the magnitude component of a 4D Flow image and (b) in an MRA image. The improved sharpness in the aorta, apparent in the MRA frame, illustrates its advantage in the extraction of accurate segmentations of the aortic anatomy.
}
\label{fig:raw_data}
\end{figure}

\begin{figure}
   \scalebox{-1}[1]{
       \includegraphics[width=.35\linewidth]{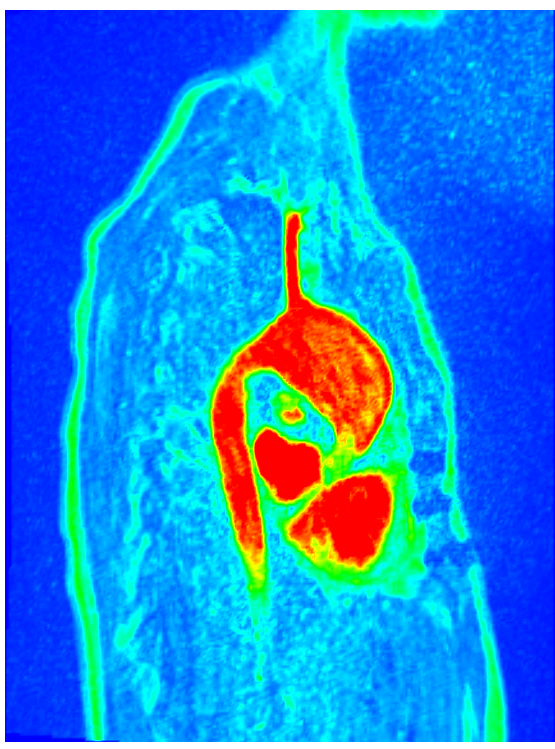} 
   }
   \includegraphics[width=.3\linewidth]{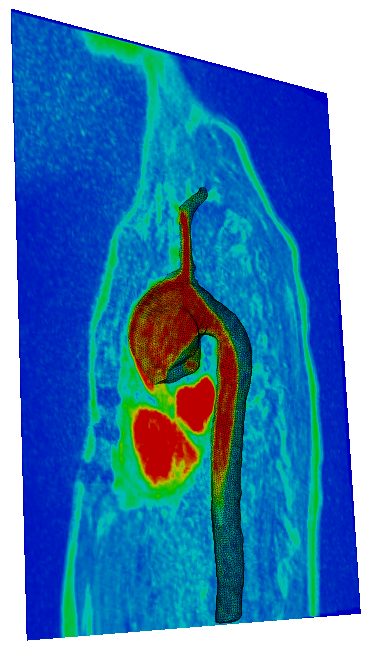}
   \includegraphics[width=.2\linewidth]{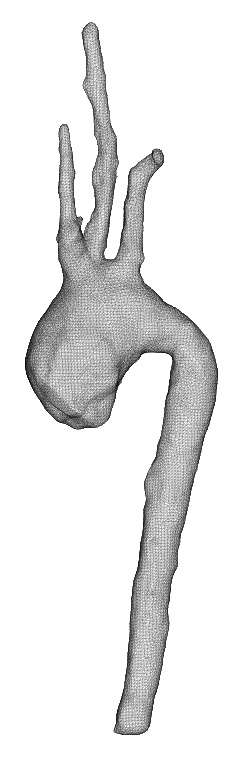}
   \caption{Segmentation of the aorta from the MRA image.}
\label{fig:mra_segmentation_overlay}
\end{figure}

Patients who undergo 4D Flow scans frequently also undergo MRA scans. MRA images usually have higher spatial resolution than 4D Flow images and do not suffer from the incompleteness and bias mentioned above. As such, MRA images admit more accurate segmentations. Refer to a comparison of 4D Flow and MRA data in Figure~\ref{fig:raw_data}. Figure \ref{fig:mra_segmentation_overlay} depicts a segmentation generated from an MRA image. A strategy to obtain more accurate estimates of blood flow properties, which is the focus of this paper, combines velocity measurements extracted from 4D Flow images with vessel segmentations extracted from MRA images. A high-level description of an implementation strategy is given in Algorithm \ref{alg:target}. 

\begin{algorithm}[!ht]
\caption{Aligning MRA and 4D Flow for the Estimation of Blood Flow Properties}
\label{alg:target}
\begin{algorithmic}[1] 
\State A segmentation of a vessel network of interest is extracted from the MRA image (see Figure \ref{fig:mra_segmentation_overlay}). 
\State A {\em target frame} of the 4D Flow image, in which the vessel network is depicted, is identified.
\State A displacement map $\Psi$ that aligns the MRA segmentation with the target frame is computed. 
\State The (single) transformed MRA segmentation is used with each frame of the 4D Flow image to estimate blood flow properties throughout the cardiac cycle.
\end{algorithmic}
\end{algorithm}

\begin{figure}[!ht]
\centering
\begin{subfigure}{0.35\textwidth}
  \includegraphics[width=1\linewidth]{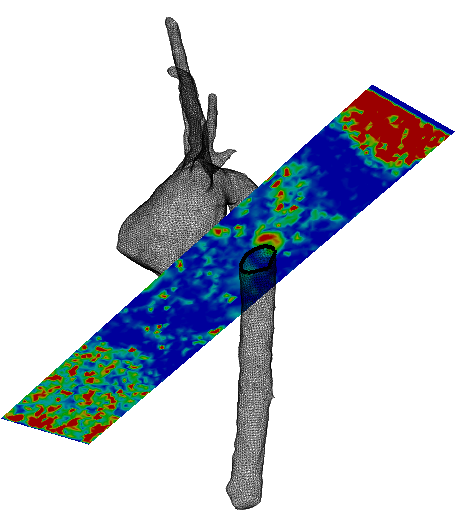}
\end{subfigure}
\begin{subfigure}{0.35\textwidth}
  \includegraphics[width=1\linewidth]{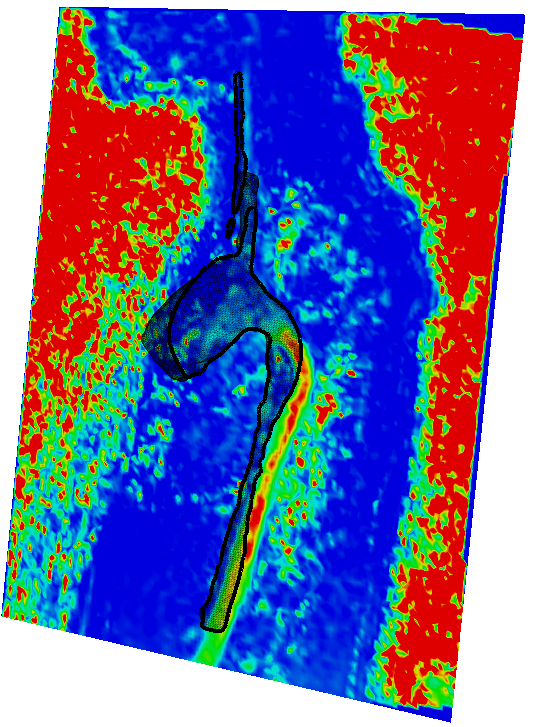}
\end{subfigure}
\centering
\begin{subfigure}{0.35\textwidth}
  \includegraphics[width=.5\linewidth]{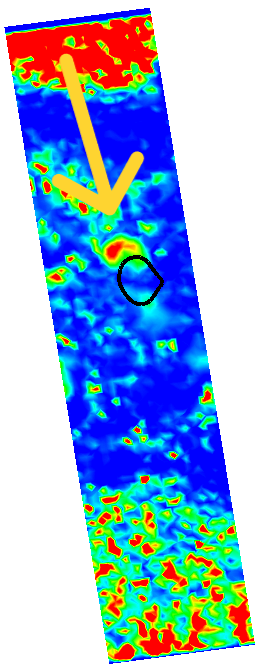}
\end{subfigure}
\begin{subfigure}{0.35\textwidth}
  \includegraphics[width=1\linewidth]{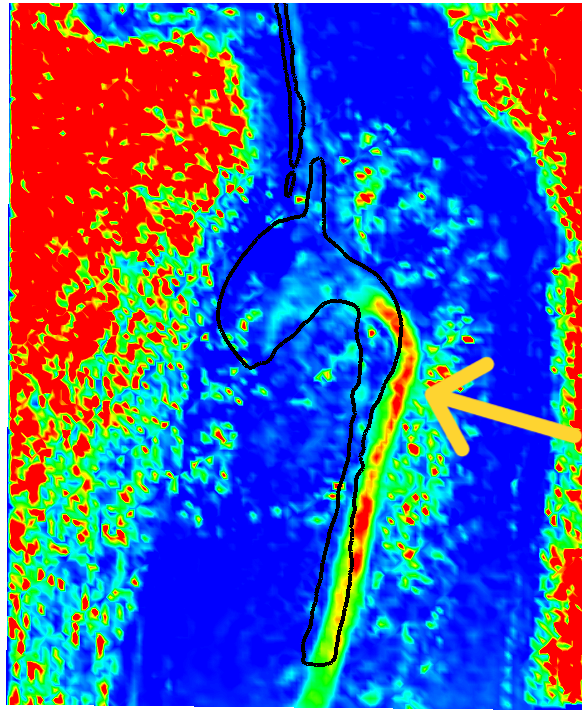}
\end{subfigure}
\caption{Overlays and their projections of the segmented aorta (from MRA) with two sections of the blood speed at systole (from 4D Flow). The arrows indicate some local regions of misalignment.}
\label{fig:misaligned_aorta}
\end{figure}

Referring to Algorithm \ref{alg:target}, the target frame of the 4D Flow image generally corresponds to a point in the systolic phase of the cardiac cycle, while the MRA segmentation corresponds to a {\em quiet} phase (characteristically, near mid-diastole, as determined during acquisition of the MRA) of the cardiac cycle. This phase discrepancy, combined with patient movement between scans, results in misalignment between the MRA and 4D Flow images. This misalignment is illustrated in Figure~\ref{fig:misaligned_aorta}. The constructed displacement map $\Psi$ is designed to correct this misalignment in order to use the aligned MRA segmentation for processing the 4D Flow image.

There are several existing approaches for registering 4D Flow images to scalar images (e.g.~MRA images. Gupta et al.~registered multi-slice balanced steady-state precession magnetic resonance images to 4D Flow images \cite{Gupta2018}). A sequence of 2D rigid transformations on the slices of the former image were computed, and an optimal representative was selected according to a cost function. Sabokrohiyeh et al.~focused on improving localization, a particular aspect of the registration algorithm from Gupta et al. \cite{Gupta2018, Sabokrohiyeh2019SketchbasedRO} While the resulting registration transformation was not necessarily rigid, it was given by a transformation matrix and hence affine. Sang et al.~presented a deep learning approach in which a motion representation model was trained using 10 multi-detector computed tomography coronary angiography images \cite{Sang2022, sang2021}. Bustamente et al.~used an atlas to automatically segment 4D Flow images \cite{Bustamante2015}. Registration was a part of the underlying construction, but registration of MRI to 4D Flow images was not in the scope of their paper.


The principal novelty of the algorithm is that, in recognizing the unreliability of geometry information extracted from 4D Flow images, it does not use segmentations extracted from 4D Flow images to perform the alignment. In particular, our approach is \textit{not} a surface registration algorithm, although it can be used to align vessel surfaces in the two images. Other distinguishing properties of the our algorithm are that it does not use machine learning techniques, and it does not restrict the aligning transformation to be rigid or even affine. Furthermore, our approach is not restricted to MRA and 4D Flow images; it can be used to align any set of images that contain ``vessel-like'' structures that admit centerline networks.

A brief summary of the algorithm is as follows. For a vessel network of interest, represented in both images, the centerlines of individual vessels are extracted. A curve registration algorithm is used to simultaneously register all centerlines extracted from the one image with their corresponding centerlines extracted from the other image. A smooth displacement map $\Psi$, represented by a Cartesian grid of B-spline functions, is calibrated to approximately extend the centerline registration transformation to the entire imaged region. The aligned MRA image is given by the composition of $\Psi$ with the original MRA image. The resulting registration algorithm is feature-based, multimodal, and deformable.

Unlike the approaches of Gupta et al.\cite{Gupta2018} and Sabokrohiyeh et al. \cite{Sabokrohiyeh2019SketchbasedRO}, the registration transformation (i.e.~the displacement field) is not restricted to be an affine map, and in particular not a rigid map. Modeling motion of internal organs with affinities is quite restrictive. The larger class of possible transformations obtained by removing this restriction affords a more accurate registration. Unlike the machine learning approaches of Sang et al.\cite{Sang2022}, the current method is not limited by the relatively small class of images used to train the underlying neural network. This feature is of particular importance for our intended application, namely to characterize hemodynamics in patients with congenital heart disease. Given the rarity of these diseases, the plausible training sets are small in sample size, and furthermore, they contain highly heterogeneous anatomy. These properties render the application of machine learning techniques more difficult.

The MRA and 4D Flow images used to illustrate and evaluate the methods in this paper come from patients with a Fontan circulation. Data was acquired at Texas Children's Hospital under protocol H-46224: ``Four-Dimensional Flow Cardiovascular Magnetic Resonance for the Assessment of Aortic Arch Properties in Single Ventricle Patients'' approved by the Baylor College of Medicine Institutional Review Board.  Throughout the paper, constructions are illustrated with a running example based on the images from a one patient in this cohort.

The remainder of the paper is organized as follows.  Section \ref{sec:methods} contains a description of the construction of the displacement map $\Psi$ and several quantitative measures of the resulting alignment. In Section \ref{sec:results}, illustrations of the resulting alignment for the running example are given. Also, in that section, distance histograms and graphs of computed blood flow (before and after alignment) are presented for all seven patient data sets. Section \ref{sec:discussion} includes a discussion of the limitations of the current method and plans for future work. Finally, in Section \ref{sec:terminology}, definitions are provided for various technical terms used in the paper. Some of the terminology is particular to this paper and some of the definitions deviate slightly from common usage to better indicate how existing terms should be understood in the present context.

\section{Methods} \label{sec:methods}

The main goal of this section is a description of the construction of the displacement map $\Psi$. Given a segmentation of the MRA image and a target frame of the 4D Flow image, the construction of the displacement map $\Psi$ is designed to reduce the distance (see Section \ref{sec:alignment_evaluation}) between the $\Psi$-transformed segmentation and the target frame. A high-level description of the procedure to construct the displacement map $\Psi$ is given in Algorithm \ref{alg:cap}. The following subsections contain detailed descriptions of individual aspects of the method. 

\begin{algorithm}[!ht]
\caption{Construction of the Displacement Map $\Psi$}\label{alg:cap}
\begin{algorithmic}[1] 
\State Composite frames are created from the 4D Flow and MRA images. 
\State A segmentation of the vessel network of interest is extracted from the composite MRA image. 
\State Centerline networks of the vessel network of interest are extracted from the MRA segmentation and from the composite frame of the 4D Flow image. 
\State Centerline networks are registered using the polyline registration algorithm (Section \ref{sec:curve_registration}).
\State The transformation of centerline networks produced in the previous step is extended to the imaged region, denoted by the map $\Psi$, using a grid of B-splines (Section \ref{sec:extension}).
\end{algorithmic}
\end{algorithm}

\subsection{Extraction of Vessel Surfaces and Centerlines}

For both MRA and 4D Flow images, different vessels are more clearly depicted in different frames. For example, the frames of the 4D Flow image corresponding to the systolic phase of the cardiac cycle highlight the aorta, while frames later in the cycle better highlight the pulmonary arteries. In the case of the MRA image, different frames correspond to different elapsed times from initial injection of a contrast agent. As such, vessels immediately downstream of the injection site are more apparent in earlier frames than the vessels further downstream. 

It is also important to mention that vessels represented in the 4D Flow image are entirely represented in the corresponding MRA image, while corresponding vessels represented in the MRA image are often only partially detectable or completely undetectable in the 4D Flow image.  As a result, the centerlines of 4D Flow networks tend to be shorter than their corresponding centerlines in the MRA networks. Considering the inherent asymmetry of the polyline registration algorithms regarding the roles of the fixed versus floating objects, it is preferable that the centerline network extracted from the MRA image assumes the role of the {\em fixed} object and that the centerline network extracted from 4D Flow image assume the role of the {\em floating} object. 

\subsubsection{Composite Images and Segmentations}
Upon manual inspection, a selection is made from the frames of the MRA image. The selection includes frames in which the vessels of interest are prominently depicted. The selected frames are combined (averaged) into a single-frame \emph{composite} scalar image. Similarly, a single-frame composite image is constructed from the 4D Flow image.
A triangular mesh representing the vessel network is segmented from the MRA composite image. The segmentations used to generate the results presented in this paper were obtained using freely available segmentation software, 3D Slicer. 

\subsubsection{Centerlines}

\begin{figure}[!ht]
\centering
\begin{subfigure}{0.2\textwidth}
  \includegraphics[width=.7\linewidth]{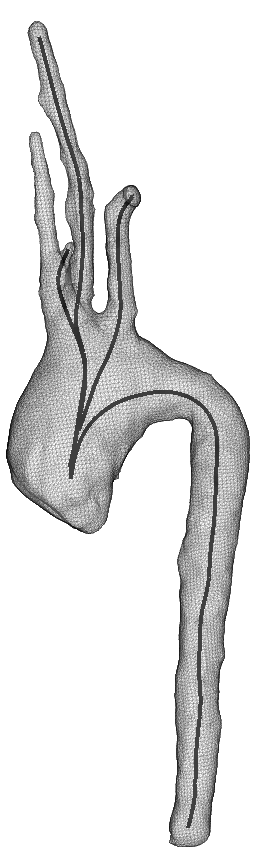}
\caption{}
\end{subfigure}
\begin{subfigure}{0.39\textwidth}
  \includegraphics[width=1\linewidth]{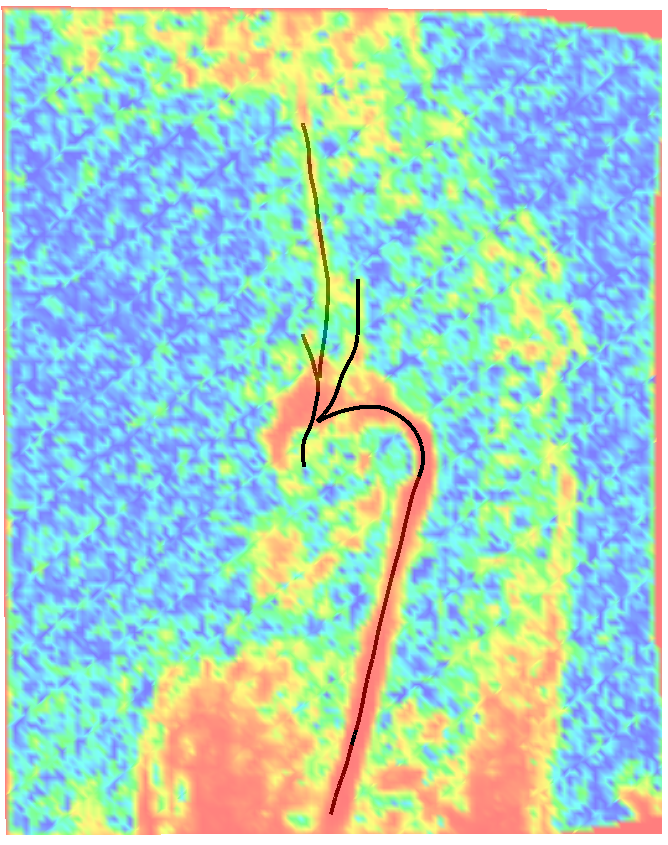}
  \caption{}
\end{subfigure}
\caption{Centerlines extracted from the segmentation of (a) an MRA image and (b) 
the coherence of the corresponding 4D Flow image.}
\label{fig:surfaces_and_centerlines}
\end{figure}

Vessels depicted in both the MRA and 4D Flow images are identified. Centerlines are extracted from the MRA segmentation and directly from a 4D Flow composite image created from a {\em coherence} field. Refer to Appendix \ref{sec:coherence} for a description of the process for creating this scalar coherence field from the velocity vectors contained in the 4D Flow images. Figure~\ref{fig:surfaces_and_centerlines} depicts the centerline networks extracted from both the MRA segmentation and 4D Flow coherence field.

\subsection{Registration of Polylines and Polyline Networks}

Centerlines extracted from medical images are represented by continuous, piecewise flat curves called \emph{polylines}. Polylines are naturally represented by computers and can be constructed to approximate arbitrary curves (e.g.~centerlines) with a simple sampling procedure. The registration procedures described in this section apply to centerlines, but are stated for polylines to emphasize their more general applicability.

\subsubsection{Rigid Registration of Polylines}

The algorithm used here for the rigid registration of polylines is a variant of the well-known iterated closest point algorithm \cite{Besl92}. At each iteration, an incremental rigid transformation $dL$ is constructed and applied to the ``current" floating polyline. The transformation $dL$ is selected to bring the current floating polyline closer to the fixed polyline (see Algorithm \ref{alg:incremental_rigid_transformation}). The algorithm halts on convergence or when the number of iterations exceeds a prescribed value. Convergence is determined by condition:
\[
    \| Id-dL \| _{F} < \mu
\]
where $\mu$ is a specified tolerance, $Id$ is the identity transformation, and $\|\cdot\|_F$ is the Frobenius norm. The final rigid transformation, denoted $L$, is the composition of all incremental rigid transformations $dL$.

\begin{algorithm}
\caption{Computing the Incremental Rigid Transformation $dL$}\label{alg:incremental_rigid_transformation}
\begin{algorithmic}[1] 
\State The closest step $\sigma$ and the closest projection $\pi$ from the floating polyline to the fixed polyline are computed.
\State For fixed positive tolerances $\epsilon_1$, $\epsilon_2$, $\epsilon_3$ (specified as parameters of the algorithm) the domains of $\sigma$ and $\pi$ are restricted to a subset $I$ of the vertices of the current floating polyline. The elements $\mathbf{u} \in I$ are selected so that:
    \begin{itemize}
        \item The angle between the tangent line of the floating polyline at $\mathbf{u}$ that of the fixed polyline at $\sigma(\mathbf{u})$ is bounded by a threshold $\epsilon_1$.
        \item The absolute difference between the curvature of the floating polyline at $\mathbf{u}$ and that of the fixed polyline at $\sigma(\mathbf{u})$ is bounded by a threshold $\epsilon_2$.
        \item The absolute difference between the torsion of the floating polyline at $\mathbf{u}$ and that of the fixed polyline at $\sigma(\mathbf{u})$ is bounded by a threshold $\epsilon_3$.
    \end{itemize}
    (Note that the image of the floating polyline is not expected to match the fixed polyline in its entirety. Restricting the domain of $\sigma$ and $\pi$ to $I$ localizes the incremental matching procedure).
\State The rigid transformation, $dL$, that maps the points $\mathbf{u} \in I$ as close as possible to their corresponding points $\pi(\mathbf{u})$, is computed. As long as the points $\mathbf{u} \in I$ are not collinear, the transformation is unique and can be readily computed as the solution of the well-known ``Procrustes problem."
\end{algorithmic}
\end{algorithm}

\subsubsection{Nonrigid Registration of Polylines}

As with the rigid registration presented above, the nonrigid transformation of the floating polyline, denoted $X_\text{floating}$, is a composition of incremental transformations that bring the floating polyline progressively closer to the fixed polyline, denoted $X_\text{fixed}$. A mode parameter specifies one of three available variants of the incremental transformation. The stopping criteria of the iterative process includes a maximum iteration limit and a convergence criterion determined by this mode. In all modes, the procedure for generating each incremental nonrigid transformation $dT$ begins with the computation of the closest step $\sigma$ and closest projection $\pi$. Subsequent steps of the incremental procedure depend both on the mode and on parameters  $\epsilon > 0$ and $0 < \mu < 1$. The parameter $\mu$ is akin to a step size; higher values increase the speed of convergence but also reduce the chance of successful convergence. The parameter $\epsilon$ is a threshold for determining convergence of the iterative process. The reported nonrigid transformation, denoted $T$, is the composition of all incremental transformations $dT$.

A brief description of the modes is included here for completeness. In {\em Mode 1}, each of the vertices $\mathbf{u}$ of $X$ is moved uniformly towards their closest projection $\pi(\mathbf{u})$ as follows:
\begin{align}
dT(\mathbf{u}) := (1-\mu)\mathbf{u} + \mu\, \pi(\mathbf{u})
\end{align}
until a convergence criterion is satisfied:
\begin{align}
 \max_\mathbf{u} ||dT(\mathbf{u}) - \pi(\mathbf{u})|| < \epsilon   
\end{align}           
In {\em Mode 2}, all the vertices of $X$ are moved simultaneously (not necessarily uniformly) towards their closest projection so that their tangent directions approach those of their closest projections, and so that their centroid approaches the centroid of their closest projections. In this mode, $dT$ is determined implicitly by the relations:
\begin{align}
 \theta(dT(\mathbf{u}), \sigma(\mathbf{u})) = \mu \theta(\mathbf{u}, \sigma(\mathbf{u})) \quad\quad |\mathbf{c} - \mathbf{c}_2| = \mu |\mathbf{c}_1 - \mathbf{c}_2|          
\end{align}
with convergence criteria:
\begin{align}
\max_{\mathbf{u}}|\theta(\mathbf{u}, \sigma(\mathbf{u}))| < \epsilon \quad\quad |\mathbf{c} - \mathbf{c}_2| < \epsilon
\end{align}
Here, $\theta(\mathbf{p},\mathbf{q})$ denotes the angle between the tangent lines at $\mathbf{p}$ and $\mathbf{q}$, and $\mathbf{c}$, $\mathbf{c}_1$, $\mathbf{c}_2$ denote, respectively, the centroids of the polylines $dT(X)$, $\pi(X)$ and $X$.
Finally, {\em Mode 3} consists of simultaneously moving all  vertices of $X$ (not necessarily uniformly) towards their closest projection so that $(i)$ their curvatures and their torsions approach those of their closest projections, $(ii)$ their centroid uniformly approaches the centroid of their closest projections and $(iii)$ their principal component directions approach those of their closest projections. The first mode is used exclusively for the examples in this paper, but more complex data sets are sometimes encountered, whose analysis benefits from the last two modes. These modes are more sophisticated, both conceptually and computationally, than the first. 

\subsubsection{Polyline Networks and their Registration}\label{sec:curve_registration}

The algorithm for the registration of polyline networks is essentially the simultaneous application of the rigid and deformable registration algorithms to the constituent polylines of the network. The closest step transformation of polylines can be extended \emph{blindly} or \emph{informedly} to the context of polyline networks. The  \emph{blind} closest step transformation is a direct analogue of the closest step transformation in single polyline case. Namely, for a vertex $\mathbf{u}$ of the floating polyline network, $\sigma(\mathbf{u})$ is taken to be the vertex in the fixed polyline network closest to $\mathbf{u}$. To define the \emph{informed} closest step transformation, a correspondence:
\[
    g : \{1,2,\ldots,m\} \to \{1,2,\ldots,n\}
\]
is provided between polylines $(c_1, \ldots, c_m)$ of the floating network and polylines $(d_1, \ldots, d_n)$ of the fixed network. Then, for a vertex $\mathbf{u}$ of polyline $c_i$, $\sigma(\mathbf{u})$ is taken to be the closest point to $\mathbf{u}$ on the polyline $d_{g(i)}$. The informed version can be used when it is known which polylines correspond to which vessels in both networks, which is generally the case in practice.

\begin{figure}[!ht]
\centering
\begin{subfigure}{0.2\textwidth}
  \includegraphics[width=1\linewidth]{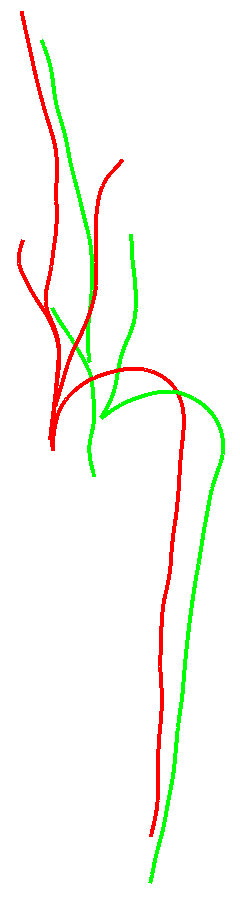}
\end{subfigure}
\begin{subfigure}{0.2\textwidth}
  \includegraphics[width=1\linewidth]{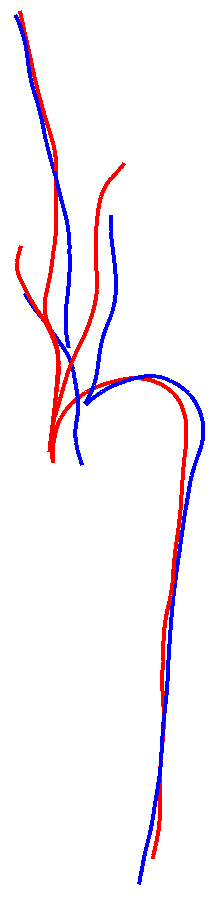}
\end{subfigure}
\begin{subfigure}{0.2\textwidth}
  \includegraphics[width=1\linewidth]{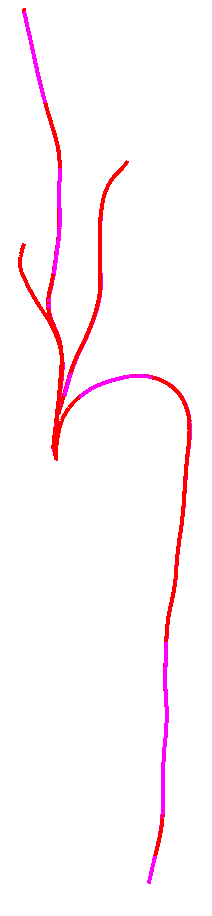}
\end{subfigure}

\caption{Polyline networks $X_\text{fixed}$ in red, $X_\text{floating}$ in green,  $L(X_\text{floating})$ in blue, and $(T\circ L)(X_\text{floating})$ in magenta.}
\label{fig:polyline_network_registration}
\end{figure}

The overall polyline network registration algorithm begins with the computation of the closest step and closest projection transformations (either in blind or informed mode) for a given pair $X_\text{floating}, X_\text{fixed}$ of polyline networks. Next, a rigid polyline transformation $L$ is generated by the rigid registration of $X_\text{floating}$ to $X_\text{fixed}$. Finally, a polyline transformation $T$ is generated by the nonrigid registration of $L(X_\text{floating})$ to $X_\text{fixed}$.  The final registered polyline network is given by the composition $T \circ L$.
If the polyline network registration is successful, then, for each vertex $\mathbf{q}$ of $L(X_\text{floating})$, $T(\mathbf{q})$ lies near a point on the fixed centerline network corresponding to the vertex $L^{-1}(\mathbf{q})$ on the floating centerline network (see Figure \ref{fig:polyline_network_registration}).


\subsubsection{Extension of the Polyline Displacement Field to the Imaged Region} \label{sec:extension}

\begin{figure*}[ht]
\centering
\includegraphics[width=0.3\textwidth]{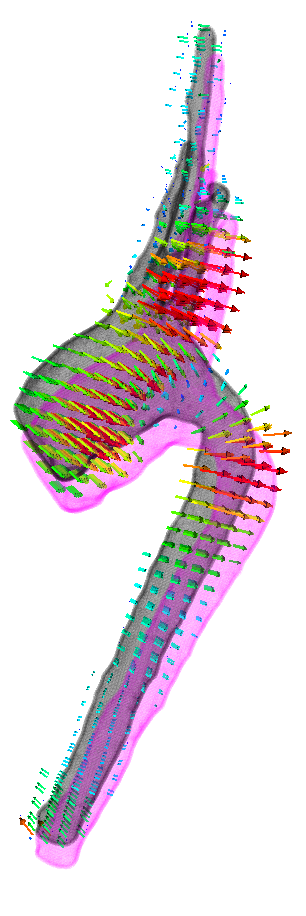}
\caption{The unaligned MRA segmentation in grey, the displacement field $\Psi$ represented by the coloured arrows, and the resulting aligned segmentation in magenta. }
\label{fig:displacements}
\end{figure*}

The centerline network registration is a determined by $T \circ L$, where $L$ is a rigid transformation and $T$ is a transformation of polyline networks that is not generally rigid. The extension of the rigid transformation $L$ to a displacement map $\widehat{L}$, defined on the entire imaged region, is uniquely determined by $L$. On the other hand, an extension of the nonrigid transformation $T$ to a displacement map $\widehat{T}$ on the imaged region is far from unique. A procedure for computing 
such an approximate extension $\widehat{T}$ is given below. The desired displacement field $\Psi$, defined on the entire imaged region, is then given by the composition $\Psi = \widehat{T} \circ \widehat{L}$. Refer to Figure~\ref{fig:displacements} for a visualization of this displacement field.
 
The displacement field $\widehat{T}$ is specified by a Cartesian grid of B-spline functions, each weighted by three coefficients (see Section \ref{sec:bspline}). The coefficients $\left\{\alpha_i, \beta_i, \gamma_i \right\}$ of the B-spline function at the $i^{th}$ grid point are computed as the solution, $\mathbf{x}$, to a linear system:

\begin{equation}
    (\mathbf{A}^T\mathbf{A} + \lambda_1 \mathbf{H}^T\mathbf{H} + \lambda_2 \mathbf{G}^T\mathbf{G})\mathbf{x} = \mathbf{A}^T\mathbf{b}
    \label{eqn:optimization}
\end{equation}
\noindent
where $\mathbf{A}$ and $\mathbf{b}$ are chosen so that the linear system $\mathbf{A} \mathbf{x} = \mathbf{b}$ expresses that $\widehat{T}$ extends the deformable registration. More explicitly, for each vertex $\mathbf{p} \in L(X_\text{floating})$:
     \begin{equation}
     \widehat{T}(\mathbf{p}) = T (\mathbf{p})
         \label{eqn:extension}
     \end{equation}
The matrix $\mathbf{H}$ is the {\em graph Laplacian}\cite{Godsil2001-in} of the Cartesian grid representation of the vessel lumen. The linear system $\mathbf{H} \mathbf{x} = \mathbf{0}$ expresses that for each vertex $\mathbf{p}$ in the vessel lumen, $widehat{T}(\mathbf{p})$ is the average value of $\widehat{T}(\mathbf{q})$ among the neighboring vertices $\mathbf{q}$ of $\mathbf{p}$. Intuitively, this condition limits the strain imposed by $\widehat{T}$ on the vessel lumen. The matrix $\mathbf{G}$ is a \emph{Tikhonov} matrix (optionally the identity matrix), and the linear system $\mathbf{G} \mathbf{x} = \mathbf{0}$ expresses that $\widehat{T} \equiv \mathbf{0}$. Intuitively, this constrains the displacement imposed by $\widehat{T}$ to be as small as possible.  


Equation \eqref{eqn:optimization} can be derived from an over-constrained least-squares problem with regularization terms defined by the matrices $\mathbf{H}$ and $\mathbf{G}$. The objectives specified above are not simultaneously satisfiable, but their relative importance in the corresponding optimization problem are prescribed by the regularization parameters $\lambda_1$ and $\lambda_2$.

\subsection{Evaluation of Alignment} \label{sec:alignment_evaluation}

Commonly, a registration between two surfaces is evaluated according to some measure of the distance between the two surfaces, both before and after alignment. A salient feature of the alignment approach presented in this paper is that it is {\em not} an alignment of surfaces. Indeed, surfaces extracted from 4D Flow images are deemed unreliable, as discussed above, and are not directly used in the alignment algorithm.

The extent to which a displacement map $\Psi \colon \mathbb{R}^3 \to \mathbb{R}^3$ aligns an MRA image with a 4D Flow image is captured by distance histograms between the vessel centerline networks generated from both of these modalities. More precisely, a distance histogram is created from the centerline network extracted from the MRA segmentation, transformed by $\Psi$, and the centerline network extracted from a scalar coherence image (see Section \ref{sec:coherence}) generated from the 4D Flow data. A secondary method to measure the efficacy of an alignment uses a ``monotonicity property" of the fluid flow through a vessel cross-section.
A section of the aorta unlikely to experience back flow is selected; any section of the aorta downstream of the left subclavian artery qualifies. Note that small area-preserving deformations of this section tend to result in smaller estimates of fluid flow rate. Estimates of flow rates across that section are computed using the unaligned MRA segmentation and then with the aligned MRA segmentation. The extent to which the ``aligned" estimates are larger than the ``unaligned" estimates suggests improved alignment. 



\section{Results} \label{sec:results}

\begin{figure}[!ht]
\centering

\begin{subfigure}{0.2\textwidth}
  \includegraphics[width=1\linewidth]{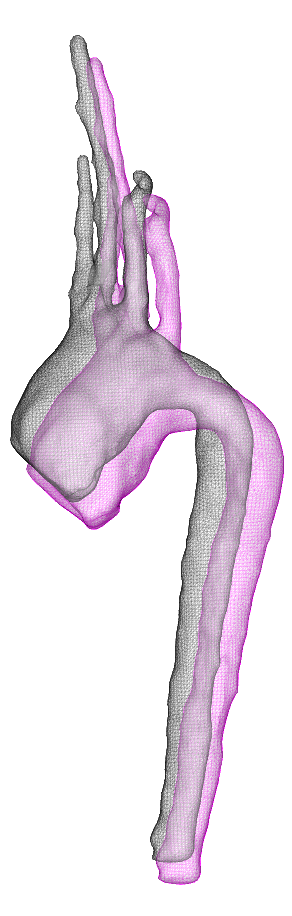}
  \caption{}
\end{subfigure}
\begin{subfigure}{0.25\textwidth}
  \includegraphics[width=1\linewidth]{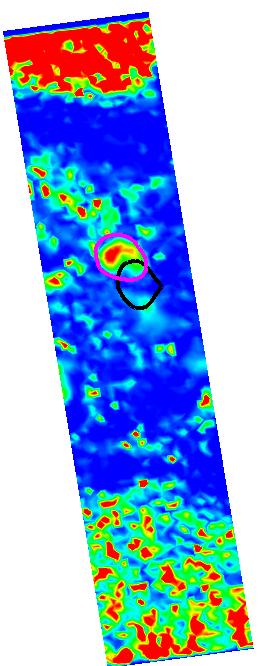}
  \caption{}
\end{subfigure}
\begin{subfigure}{0.4\textwidth}
   \scalebox{-1}[1]{
       \includegraphics[width=1\linewidth]{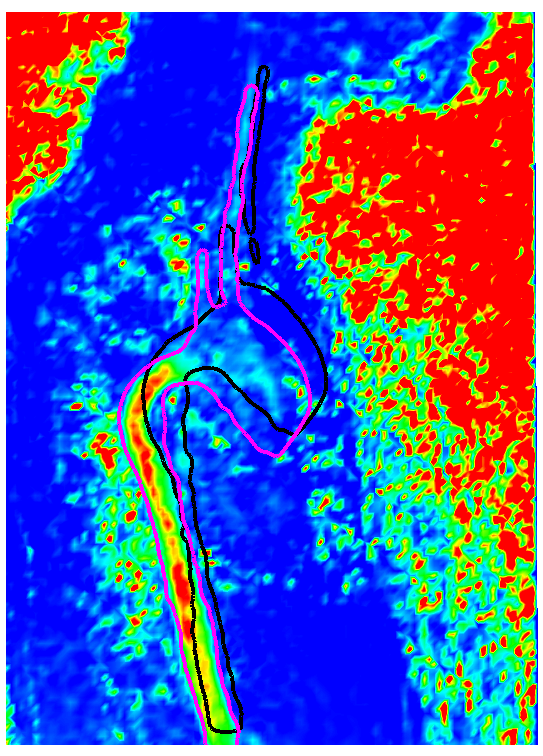} 
   }
  \caption{}
\end{subfigure}

\caption{Panel (a) shows the aligned MRA (AMRA) segmentation  in magenta and the   unaligned MRA (UMRA) segmentation in grey. Panels (b) and (c) show two sections of these segmentations overlaid with the magnitude of the velocity field derived from the 4D Flow image. Compare with Figure \ref{fig:misaligned_aorta}.}
\label{fig:realignment}
\end{figure}

For the running example, various views of misalignment between the MRA segmentation and the 4D Flow image before and after application of the alignment procedure appear in Figure \ref{fig:realignment} where the improved alignment is apparent. In particular, the aligned segmentation more accurately surrounds regions of high velocity corresponding to flow through the vessel lumen.

\begin{figure}[!ht]
\centering
  \includegraphics[scale=0.4,trim=0 -40 -50 25]{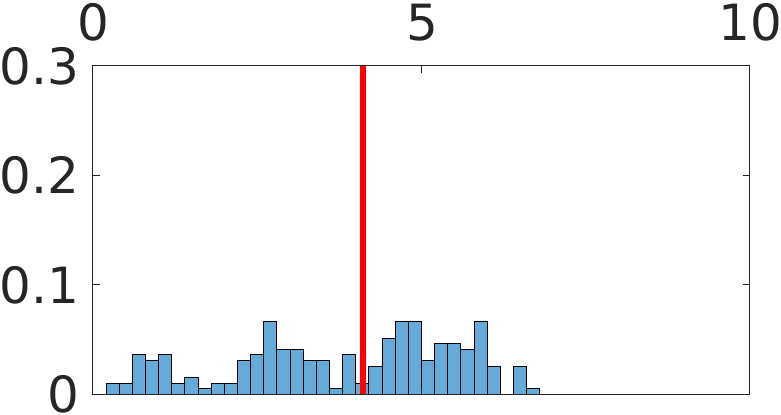}
  \put(-120, 105){distance [mm]}
  \includegraphics[scale=0.4, trim=0 -40 0 25]{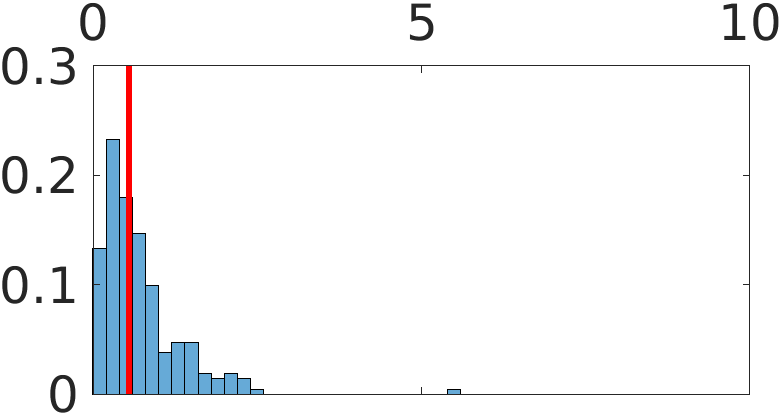}
  \put(-100, 105){distance [mm]}\\
  \includegraphics[scale=0.4, trim=-30 -40 -90 25]{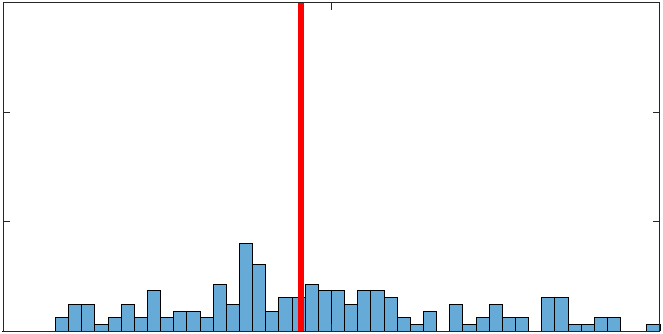}
  \includegraphics[scale=0.4,trim=-10 -40 0 25]{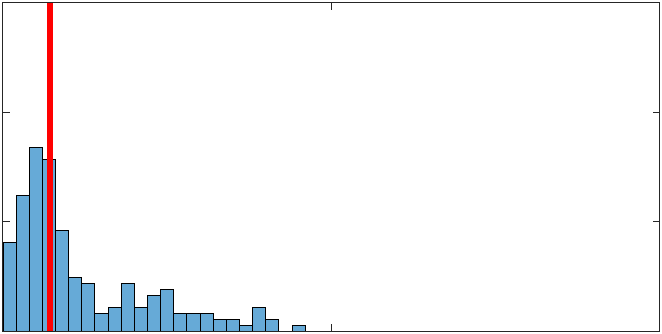} \\
  \includegraphics[scale=0.4, trim=-30 -40 -90 25]{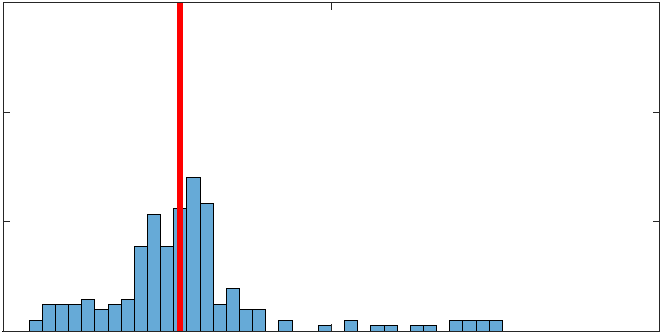}
  \includegraphics[scale=0.4,trim=-10 -40 0 25]{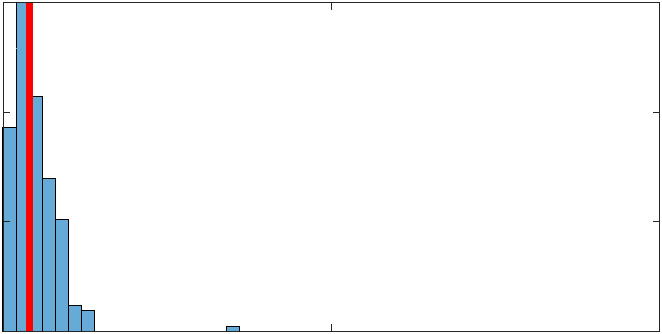} \\
  \includegraphics[scale=0.4, trim=-30 -40 -90 25]{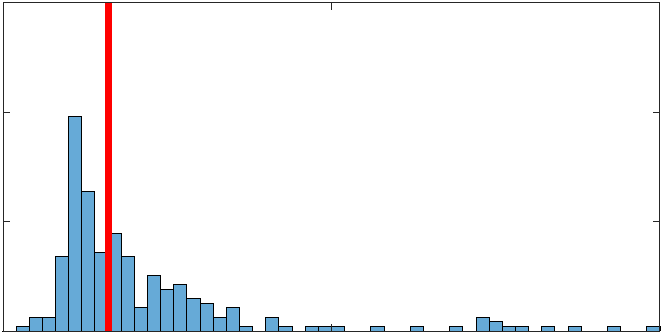}
  \includegraphics[scale=0.4,trim=-10 -40 0 25]{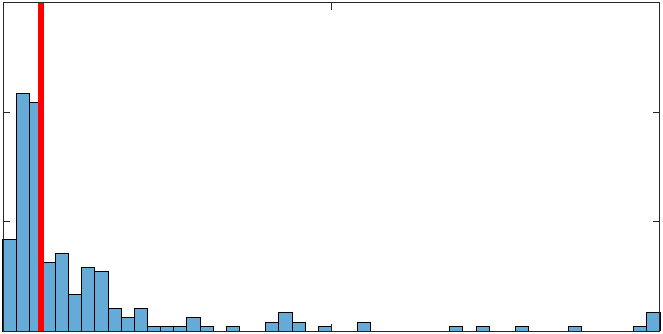} \\
  \includegraphics[scale=0.4, trim=-30 -40 -90 25]{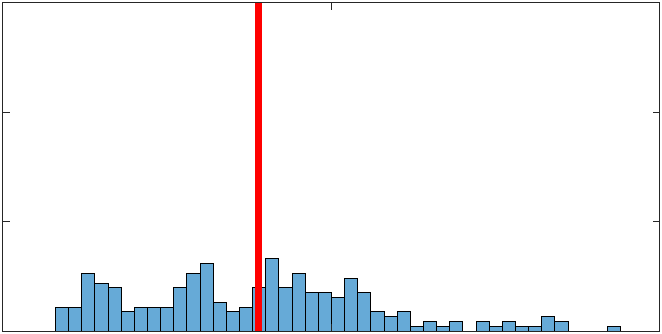}
  \includegraphics[scale=0.4,trim=-10 -40 0 25]{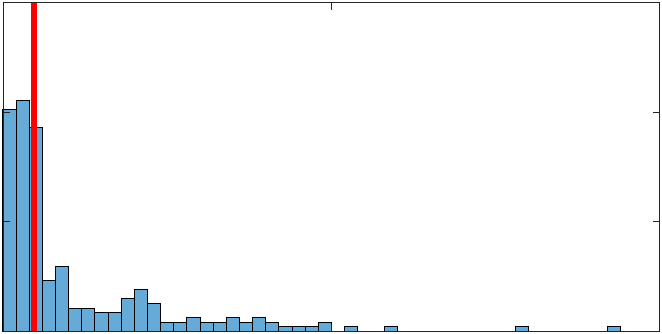} \\
  \includegraphics[scale=0.4, trim=-30 -40 -90 25]{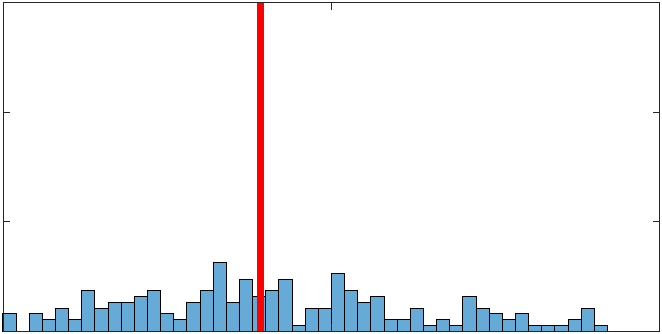}
  \includegraphics[scale=0.4,trim=-10 -40 0 25]{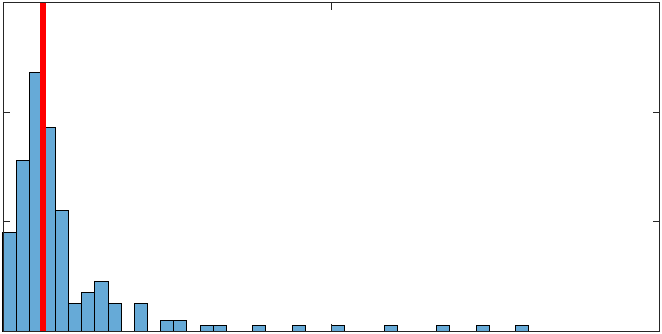} \\
  \includegraphics[scale=0.4, trim=-30 -40 -90 25]{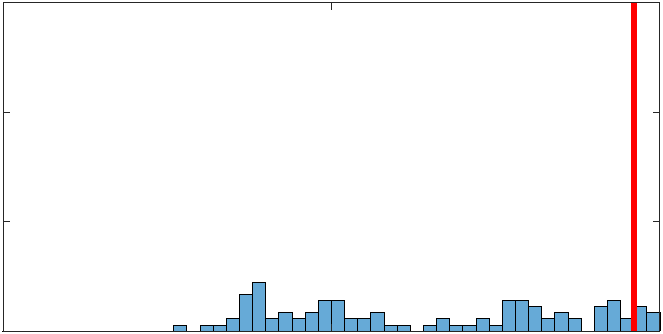}
  \includegraphics[scale=0.4,trim=-10 -40 0 25]{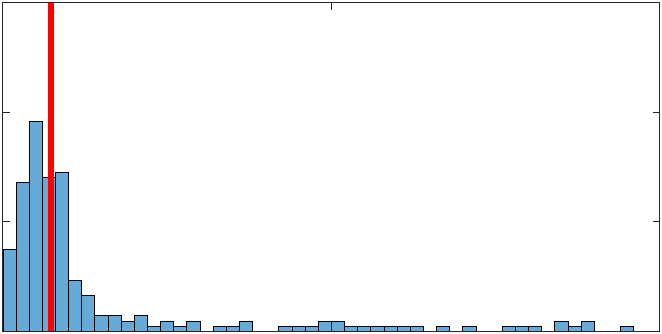} \\

  \caption{On the left (respectively right) are distance histograms between centerline networks extracted from 4D Flow coherence images and centerline networks extracted from unaligned (respectively aligned) MRA images. The red vertical lines indicate median values of the histograms.}
\label{fig:histograms}
\end{figure}

For the image pairs (MRA and 4D Flow) from the seven available data sets, fresh centerline networks (distinct from the centerlines used to perform the registration) were extracted from the 4D Flow coherence image and from both the unaligned and aligned MRA segmentations. Distance histograms for each patient appear in Figure \ref{fig:histograms}, with unaligned results in the left column and aligned results in the right column. In all cases, histograms on the right are shifted towards zero compared to their counterpart on the left. This difference indicates that the aligned MRA segmentation is closer to the 4D Flow image than the unaligned MRA segmentation. More quantitatively, for the seven analyzed patients, the relative decreases in median value of the distance histograms as a result of the alignment procedure appear in Table \ref{tab:histogram_relative_decreases}. Medians of the distance histograms decreased an average of 83.5\%.

 \begin{table}[!h]
     \centering
     \begin{tabular}{|c|c|c|}
     \hline
         $M_\text{UMRA {\text -} 4DFlow}$ (mm) & $M_\text{AMRA - 4DFlow}$ (mm) & relative decrease \\
         \hline
         4.11   & .549   & 87\%\\
         4.54   & .711   & 84\%\\
         2.69   & .406   & 85\%\\
         1.61   & .580   & 64\%\\
         3.89   & .472   & 88\%\\
         3.92   & .612   & 84\%\\
         9.61   & .733   & 92\% \\
         \hline
     \end{tabular}
     \caption{The numbers under column $M_\text{UMRA {\text -} 4DFlow}$ (resp. $M_\text{AMRA {\text -} 4DFlow}$) are the median values (in mm) of distance histograms 
     between centerline networks extracted from unaligned (resp. aligned) MRA images and those extracted from 4D Flow coherence images (see Figure \ref{fig:histograms}). The third column gives the relative decrease of these medians as a result of the alignment.} 
    
     \label{tab:histogram_relative_decreases}
 \end{table}

\begin{figure}[ht!]
\centering
  \includegraphics[scale=0.35,trim=0 0 -55 0]{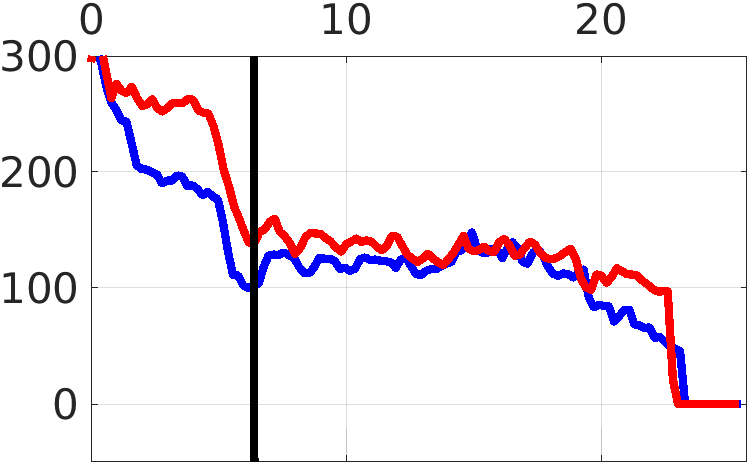}
  \put(-162, 0){\rotatebox{90}{flow rate [mL/s]}}
  \put(-110, 85){distance [cm]}
  \includegraphics[scale=0.35, trim= 0 0 0 0]{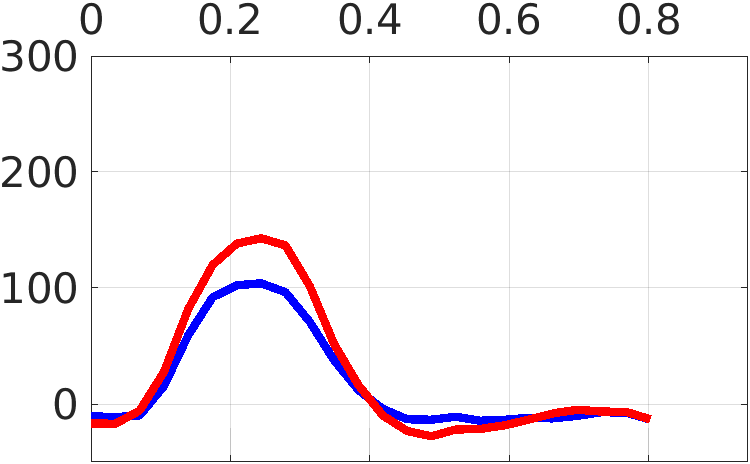}
  \put(-80, 85){time [s]}\\
  \includegraphics[scale=0.35, trim=-40 0 -90 0]{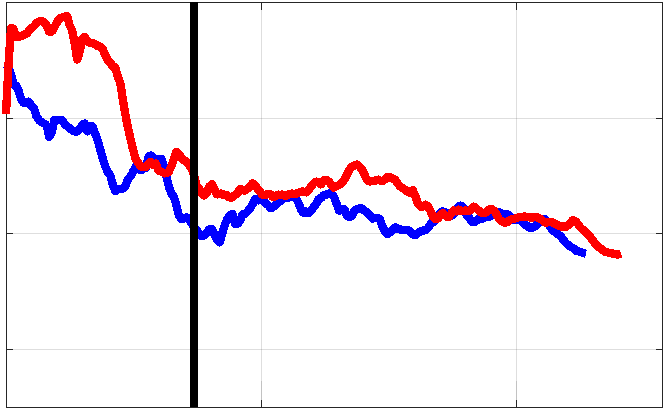}
  \includegraphics[scale=0.35,trim=0 0 0 0]{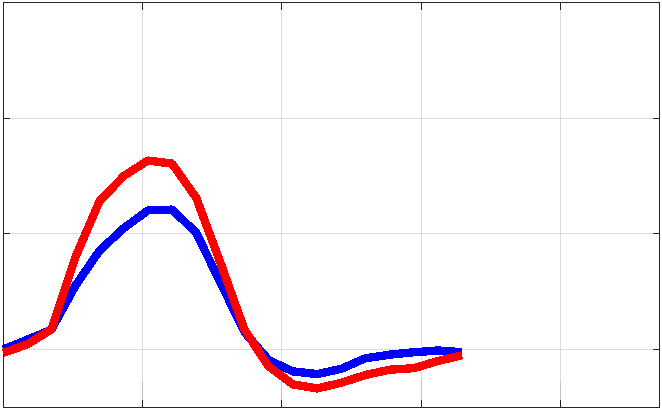} \\
  \includegraphics[scale=0.35, trim=-40 0 -90 0]{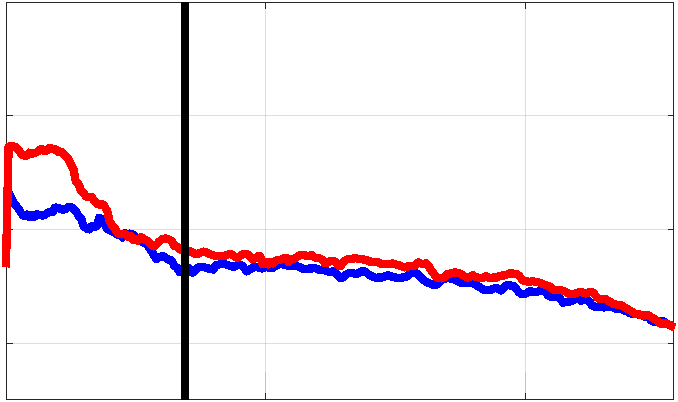}
  \includegraphics[scale=0.35,trim=0 0 0 0]{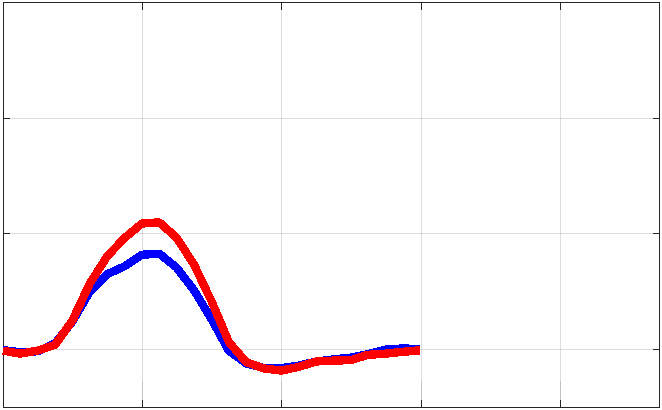} \\
  \includegraphics[scale=0.35, trim=-40 0 -90 0]{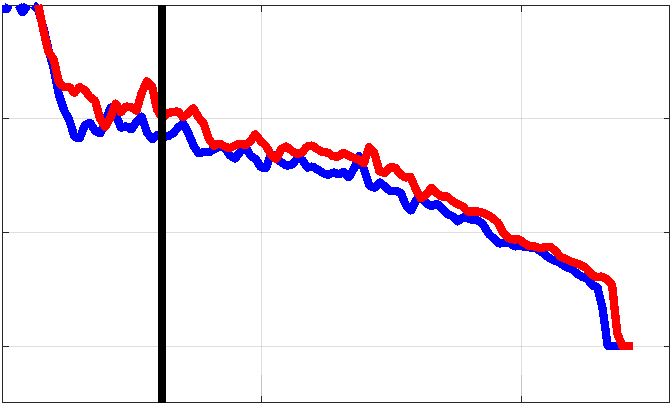}
  \includegraphics[scale=0.35,trim=0 0 0 0]{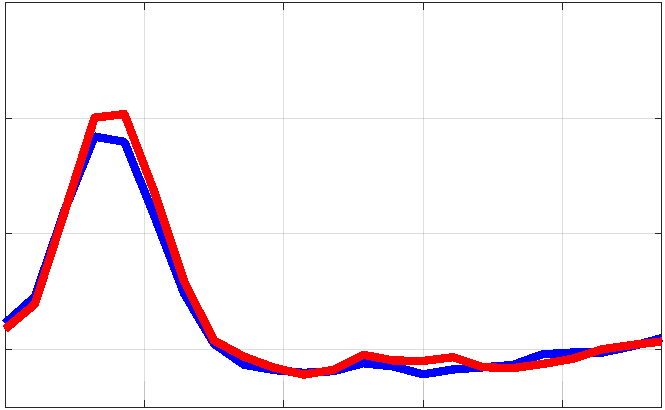} \\
  \includegraphics[scale=0.35, trim=-40 0 -90 0]{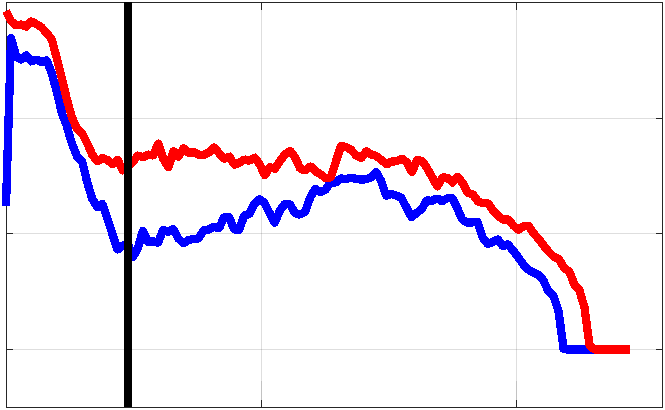}
  \includegraphics[scale=0.35,trim=0 0 0 0]{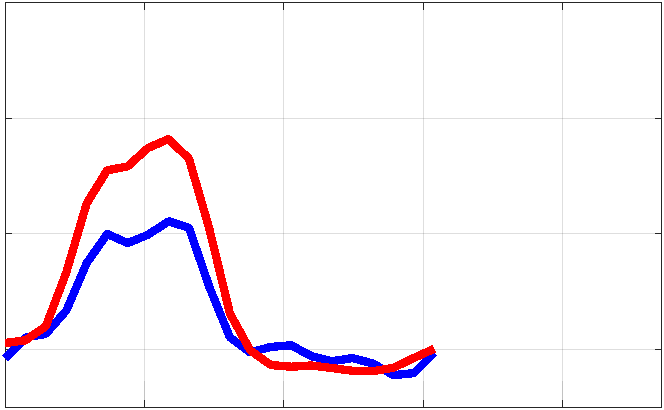} \\
  \includegraphics[scale=0.35, trim=-40 0 -90 0]{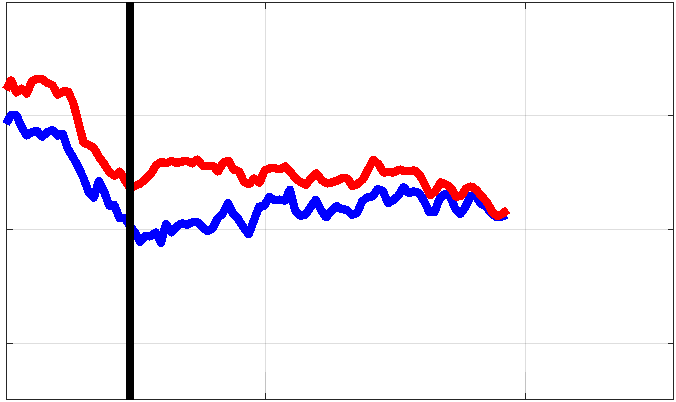}
  \includegraphics[scale=0.35,trim=0 0 0 0]{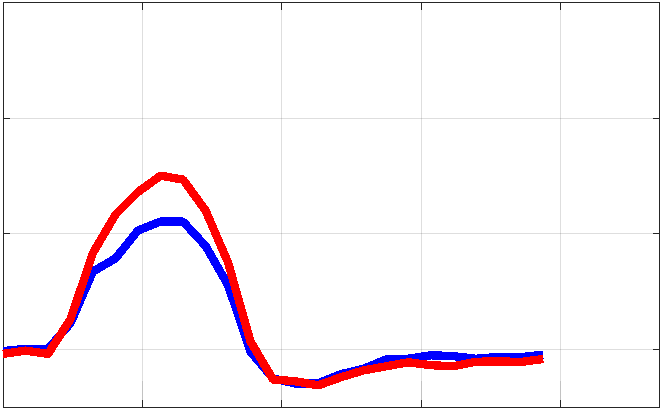} \\
  \includegraphics[scale=0.35, trim=-40 0 -90 0]{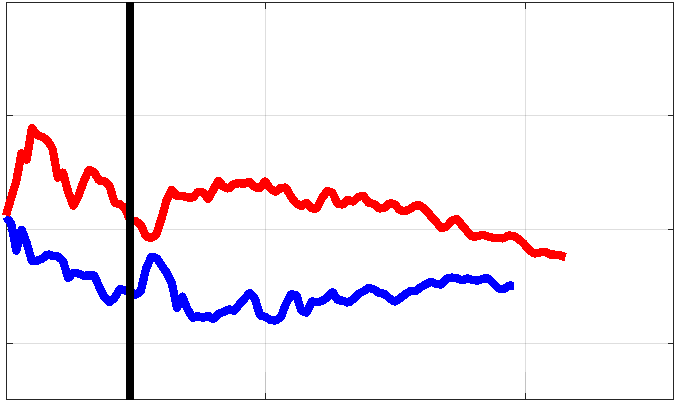}
  \includegraphics[scale=0.35,trim=0 0 0 0]{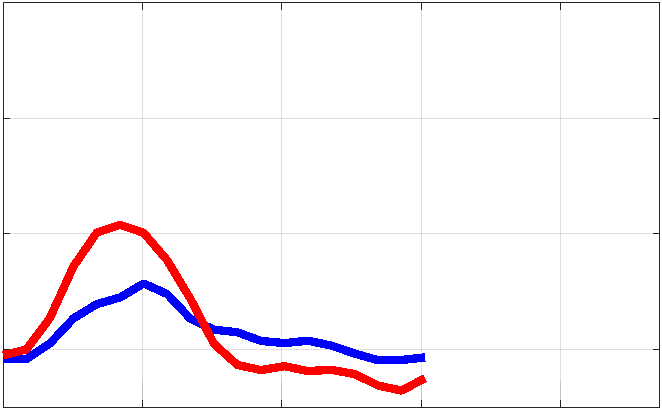} 
\caption{Graphs of blood flow computed with the unaligned (blue) and aligned (red) MRA segmentations. The left column corresponds to the flow, as a function of the centerline, measured at systole. The right column corresponds to the flow, as a function of time, measured at the location indicated by the black vertical line in the corresponding figure on the left. 
}

\label{fig:bloodflow}
\end{figure}

For the same seven patients, graphs of blood flow in the aorta computed with geometry information from the aligned MRA segmentation and velocity information from the 4D Flow image appear in Figure \ref{fig:bloodflow}. The extent to which the red line lies above the blue line reflects higher flow estimates using the aligned MRA segmentation. According to the monotonicity property, higher flow estimates suggest closer alignment of the MRA segmentation to the 4D Flow image. 

\clearpage

\clearpage


\section{Conclusions, Limitations, and Future Work} \label{sec:discussion}

This paper describes a method to align MRA and 4D Flow images. While this work was conceived to register these specific sequences, the approach is general enough to handle any modality that captures vessel structures well approximated by centerline networks. Our method relies on the direct registration of vessel centerlines, which can be readily generated from noisy image modalities such as 4D Flow. The displacement field defined on the centerline is then extended to the entire imaged region using a regularized least-squares problem. The improved alignment, as quantified by distance histograms computed for 7 patient data sets with MRA and 4D Flow data, indicate the method is effective. Blood flow estimates for these patients were greater in systole as a result of the alignment procedure. As discussed in Section \ref{sec:alignment_evaluation}, this also suggests improved alignment. 


As mentioned in Section \ref{sec:intro}, part of the motivation for using MRA images to extract vessel segmentations is the inaccuracy and bias in surfaces segmented from 4D Flow images. At first thought, it seems hypocritical to reject 4D Flow segmentations while making critical use of the images to extract centerlines. This practice is somewhat justified by the following observations. The center of a circle is independent of any nonzero scaling of the circle. Insofar as the bias towards narrower vessels is uniform along the circumference, the extraction of centerlines from vessel surfaces should not be significantly affected. Small uniformly distributed noise applied to points uniformly sampled on the circumference of a circle will have minimal effect on the center of mass of those samples. Insofar as the inaccuracy of the segmented vessel surface due to the noise and coarse spatial resolution is uniformly distributed, the extraction of centerlines from a vessel surface should not be significantly affected. 


There are limitations to our method that are worth highlighting. An MRA image provides information at a single phase of the cardiac cycle, albeit averaged over several cycles. The intended application, however, demands estimates of blood flow properties over the entire cardiac cycle. In our approach, an alignment is computed between the MRA image and a single composite image derived from the 4D Flow phases (typically several phases near mid-systole). This single registration is used to align the MRA segmentation to all other phases of the 4D Flow image.


However, the alignment could be performed separately for each frame of the 4D Flow image to more directly account for misalignment in each phase. This procedure would result in a sequence of deforming MRA segmentations over the cardiac cycle. It should be noted that vessels are often invisible in 4D Flow images at phases in which the flow is diminished (e.g.~in the ascending aorta during diastole). A practical compromise is to perform the alignment procedure at a few selected phases in which vessels of interest are depicted in the 4D Flow image. Treating these phases as representative of the entire cycle, alignments in the other phases can be interpolated (e.g.~by trigonometric interpolation) from the representative alignments. Implementing and evaluating such an approach is one of the potential extensions of this work.


Another limitation is related to the maximum iteration limit, step size, convergence tolerance, sampling distance, regularization coefficients, and parameters associated with the segmentation and centerline extraction. These were chosen empirically based on testing with the available data. A systematic study on the sensitivity of the alignment results to these parameters was not performed. Preceding further application of this algorithm to more critical analyses, such a study would be prudent. 

Another simplification of our approach is the consideration of vessel networks as collections of independent centerlines without some additional structure. In reality, these networks have a definite directed tree structure, for example, in the way that the constituent vessels/centerlines branch off of their parent vessels. This structure can inform the alignment process and potentially achieve better results.  

Finally, of the three modes described for the nonrigid polyline registration algorithm, only the simplest, mode 1, sufficed to generate the results presented in this paper. Future work aligning more complex vessel networks (e.g.~vessel networks in the lungs) will likely need to make use of the more sophisticated modes of polyline registration.

\end{spacing}

\clearpage

\section {Appendix: Terminology and Background}  \label{sec:terminology}


\subsection{Curves and Polylines}

A \emph{curve} is a smooth function $\gamma \colon [0,1] \to \mathbb{R}^n$ for some fixed positive integer $n$. A polyline is a discrete version of a curve that is used to approximate it. More precisely, a \emph{polyline} is a list of points, called \emph{vertices}, in $\mathbb{R}^n$. $\mathcal{P}^n$ denotes the set of polylines in $\mathbb{R}^n$. For a real positive number $h$, $\mathcal{P}^n_h$ denotes the set of polylines in $\mathcal{P}^n$ having constant distance $h$ between consecutive vertices. Such polylines are called \emph{uniform polylines}.

For any curve $\gamma: [0,1] \to \mathbb{R}^n$ and $h>0$, the $h$-\emph{sampled polyline} $\gamma_h \in \mathcal{P}_h^n$ \emph{approximating} $\gamma$ is given by:
\[
\gamma_h := \gamma(t_0), \ldots, \gamma(t_k)
\]
where 
\[
0 = t_0 < t_1 < \ldots < t_k \leq 1
\]
is the unique maximal set of real numbers satisfying:
\[
|\gamma(t_i) - \gamma(t_{i-1})| = h, \quad 1 \leq i \leq k
\]

A \emph{polyline network} is a list of polylines in $\mathcal{P}^n_h$ for some fixed values of $n$ and $h$. There is no assumed or required relation between the polylines of such a network. In particular, while polylines in a network may intersect (themselves or each other) the intersection points are not distinguished or specified. 

\subsubsection{Surfaces and Meshes}

Surfaces and (triangle) meshes are two-dimensional analogs of curves and polylines, respectively. The common usage of these terms is adhered to in this paper, so explicit definitions are omitted.

\subsubsection{Differential Measures of Polylines}

Differential measures of curves such as tangents, curvature and torsion are central to the motivation and implementation of various geometric algorithms on polylines, including registration algorithms. Clearly, polylines fail all but $C^0$ smoothness conditions, so they do not admit such measures\footnote{Polylines sampled from curves can inherit differential information at their vertices from their parent curve, but this information is generally invalidated when the polyline undergoes nonrigid transformation.}.  Fortunately, the literature on discrete differential geometry contains simple and efficient algorithms that compute, directly from $h$-sampled polyline approximations of curves, discrete approximations of such differential measures.\cite{Langer2005} 

\subsubsection{Transformations of Polylines}

A \emph{transformation} of a polyline is a specification of displacement for each vertex on the polyline. For example, a displacement map $\Psi \colon \mathbb{R}^3 \to \mathbb{R}^3$ determines a transformation of a polyline $\gamma = (\mathbf{p}_1, \mathbf{p}_2, \ldots, \mathbf{p}_k)$ by:
\[
    \Psi(\gamma) := (\Psi(\mathbf{p}_1), \Psi(\mathbf{p}_2), \ldots, \Psi(\mathbf{p}_k))
\]

A polyline transformation is called \emph{rigid} (resp. \emph{affine}) if it is the restriction of a rigid (resp.~affine) displacement map on the ambient Euclidean space (i.e.~$\mathbb{R}^3$). 

\begin{figure}[!ht]
\centering
\includegraphics[width=.5\textwidth]{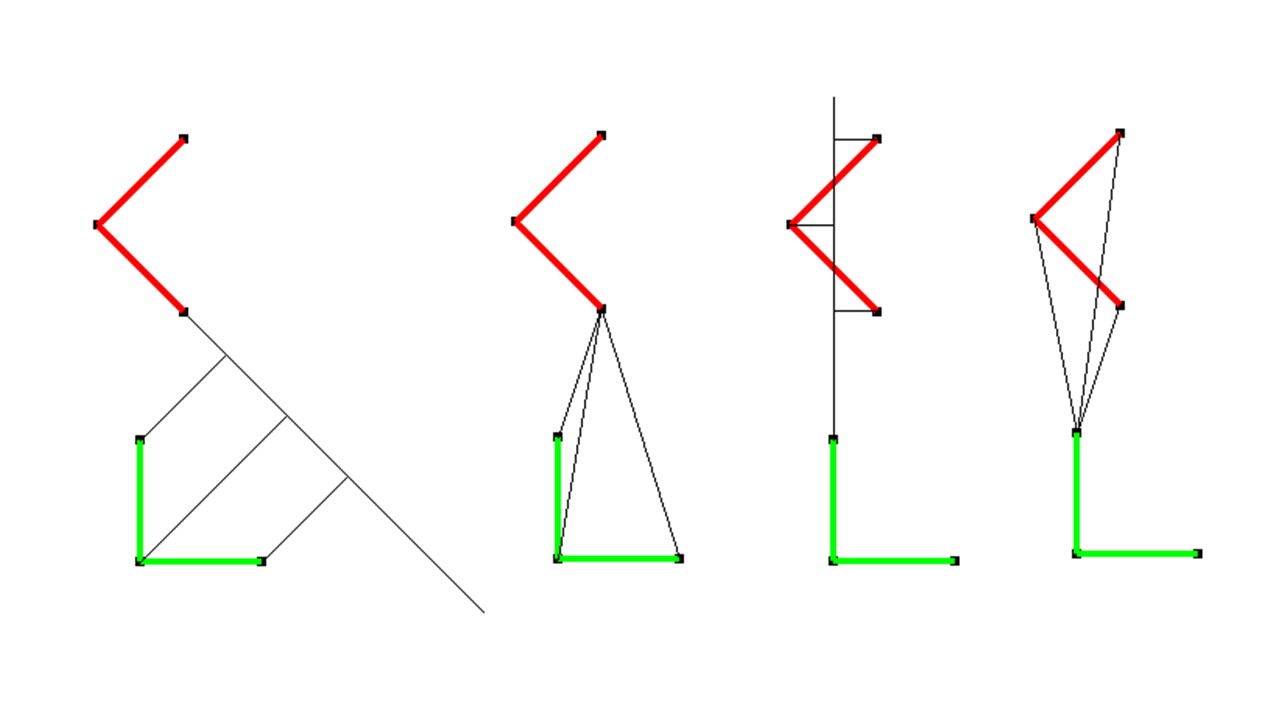}
\put(-220, 10){(a)}
\put(-140, 10){(b)}
\put(-85, 10){(c)}
\put(-40, 10){(d)}
\caption{(a) Closest projection of green to red polyline, (b) closest step of green to red polyline, (c) closest projection of red to green polyline, (d) closest step of red to green polyline.}
\label{fig:closest2}
\end{figure}

Two particular transformations of a polyline $\gamma$ are determined from a given polyline $\gamma'$. The \emph{closest step} transformation $\sigma$ sends each vertex of $\gamma$ to the closest vertex (with respect to the Euclidean distance) of $\gamma'$. The \emph{closest projection} transformation $\pi$ sends each vertex $\mathbf{p}$ of $\gamma$ to the perpendicular projection of $\mathbf{p}$ to the tangent line at $\sigma(\mathbf{p})$. Both the closest step and the closest projection are designed to move $\gamma$ closer to $\gamma'$, but the closest projection tends to preserve the length and shape of $\gamma$ better than the closest step, especially when $\gamma$ and $\gamma'$ differ significantly in length (see Figure~\ref{fig:closest2}). Note that the transformations $\sigma$ and $\pi$ are generally not affine (nor rigid). 

\subsubsection{Distance Between Polylines}\label{sec:histogram}

While they do not constitute a formal metric on $\mathcal{P}^n$, a certain pair of histograms, defined for any pair of polylines, provide an informative measure of the discrepancy between polylines.

Let $\gamma, \gamma'$ be polylines and let $\sigma$ denote the closest step transformation from $\gamma$ to $\gamma'$. For each vertex $\mathbf{w} \in \sigma(\gamma)$, and from each maximal interval in $\sigma^{-1}(\mathbf{w})$, choose a single representative. Let $V$ denote the set of all such representatives. The histogram of the set:
\[
    \left\{ ||\mathbf{v} - \pi(\mathbf{v})|| : \mathbf{v} \in V \right\},
\]
called the \emph{distance histogram} from $\gamma$ to $\gamma'$, is independent of the choices of representatives. The extent of skew towards zero of the distance histograms reflects the distance between polylines $\gamma$ and $\gamma'$, (the more skewed towards zero, the closer the polylines).  Note that the distance histogram from $\gamma$ to $\gamma'$ and from $\gamma'$ to $\gamma$ are generally not identical, but for the intended applications and for the particular patient data examined for this paper, their difference is negligible. 

\subsection{Displacement Maps}

A \emph{displacement map}, specified as a function $\Psi \colon \mathbb{R}^n \to \mathbb{R}^n$, 
is called \emph{affine} if $\Psi(\mathbf{x}) = A\mathbf{x} + \mathbf{d}$ for some fixed $\mathbf{d} \in \mathbb{R}^n$ and $n \times n$ matrix $A$ and is called \emph{rigid} if it is affine and $A$ has unit determinant.

\subsubsection{B-spline Displacement Grids and Approximate Extensions}\label{sec:bspline}

\begin{figure}[!ht]
\centering
\includegraphics[width=.3\textwidth]{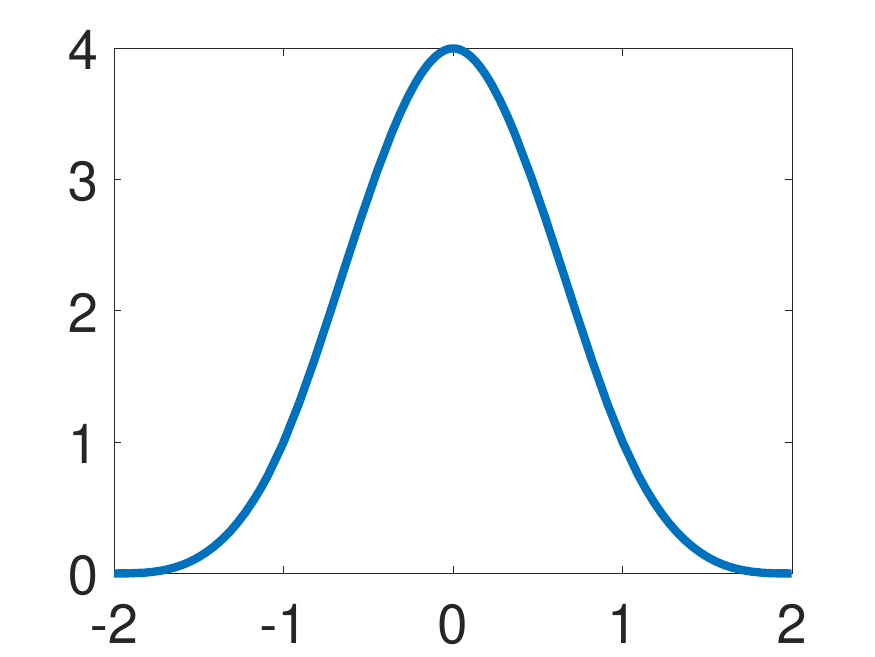}
\caption{Graph of a cubic B-Spline.}
\label{fig:bspline}
\end{figure}

A \emph{B-spline of degree $d$} refers to a certain $C^{d-1}$ function which is piecewise polynomial $\phi$ of degree $d$ and which has finite radius of support $r_s = (d+1)/2$. For example, if $d=3$, then $r_s = 2$, 
$\phi$ is defined below and the graph of $\phi(t)$ is illustrated in Figure~\ref{fig:bspline}.

\[
    \phi(t) := 
   \begin{cases}
       0, & \;\; t \leq -2 \\
       (t+r_s)^3 / 6, & -2 \leq t < -1 \\
       (-3(t+r_s)^3 + 12(t+r_s)^2 -12(t+r_s) + 4) / 6 , & -1 \leq t < 0 \\
       (3(t+r_s)^3 -24(t+r_s)^2 + 60(t+r_s) - 44) / 6 , & \;\; 0 \leq t < 1 \\
       (-(t+r_s)^3 + 12(t+r_s)^2 -48(t+r_s) + 64) / 6, & \;\; 1 \leq t < 2 \\
       0, & \;\; t  > 2 \\
   \end{cases}
\]


The function $\Phi \colon \mathbb{R}^3 \to \mathbb{R}$ defined by $\Phi(x,y,z) = \phi(x)\phi(y)\phi(z)$   
is $C^{3d-1}$, piecewise polynomial of total degree $3d$, and vanishes outside the $l_1$-ball of radius $r_s = (d+1) / 2$. 

A \emph{B-spline displacement grid} is a displacement map whose components are each linear combinations of the set $\{\Phi_i\}$ of translates of $\Phi$ by the vertices $\left\{\mathbf{p}_i\right\}$ of a rectilinear (i.e.~box-shaped) lattice in $\mathbb{R}^3$. That is, given coefficients $\alpha_i, \beta_i, \gamma_i$ for each vertex of the lattice, the displacement specified by the B-spline displacement grid at a point $\mathbf{x} \in \mathbb{R}^3$ is given by:

\begin{equation}
    \sum_i\left(\alpha_i, \beta_i, \gamma_i \right)\Phi(\mathbf{x}-\mathbf{p}_i) 
\end{equation}

Recall that a transformation of a polyline with vertex list $V$ is simply a function $T \colon V \to \mathbb{R}^3$. In this paper, an \emph{approximate extension} $\widehat{T}$ of $T$ refers to a B-spline displacement grid whose restriction to $V$ approximates $T$.

\subsection{MRA and 4D Flow Images}

An \emph{image} is a function whose domain is a rectilinear lattice in $\mathbb{R}^3$ and
whose codomain is $\mathbb{R}^n$, for some positive integer $n$. The vertices of the lattice are also referred to as \emph{voxels}. If $n = 1$, the image is called a \emph{scalar image}, otherwise, its called a \emph{vector image}. A \emph{time-resolved image} is a sequence of images varying in time but sharing a common domain and codomain. Individual images in a time-resolved image are referred to as \emph{frames}. A \emph{scan} refers to the physical process of acquiring images from a patient. 

\begin{figure}[!ht]
\centering
\begin{subfigure}{0.15\textwidth}
  \includegraphics[width=1\linewidth]{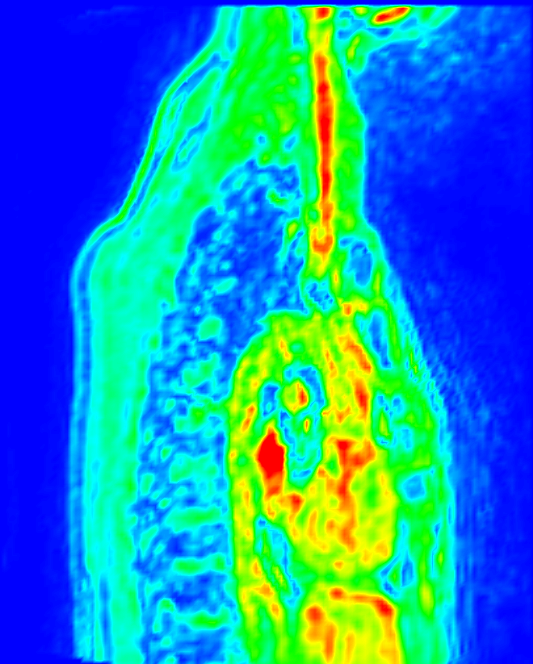}
\end{subfigure}
\begin{subfigure}{0.15\textwidth}
  \includegraphics[width=1\linewidth]{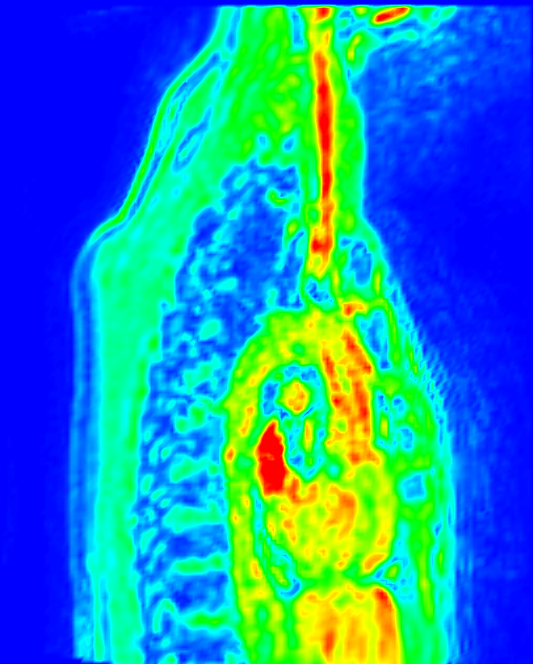}
\end{subfigure}
\begin{subfigure}{0.15\textwidth}
  \includegraphics[width=1\linewidth]{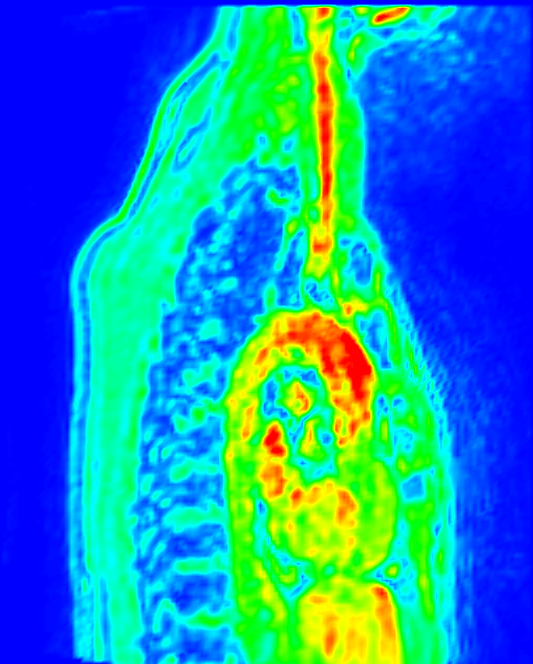}
\end{subfigure}
\begin{subfigure}{0.15\textwidth}
  \includegraphics[width=1\linewidth]{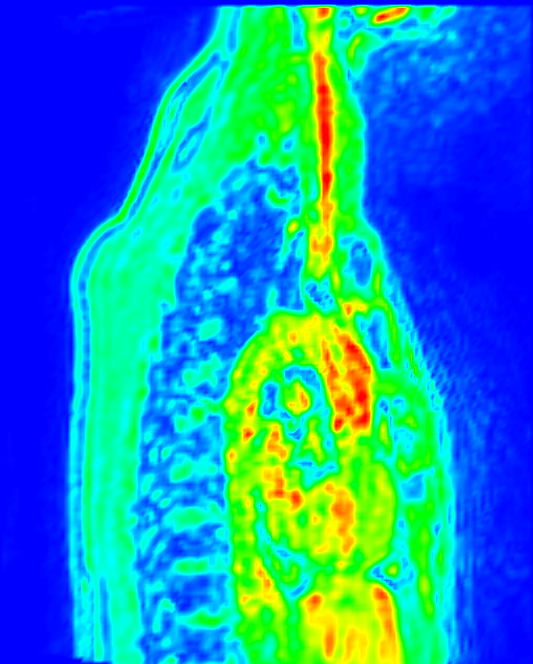}
\end{subfigure}
\begin{subfigure}{0.15\textwidth}
  \includegraphics[width=1\linewidth]{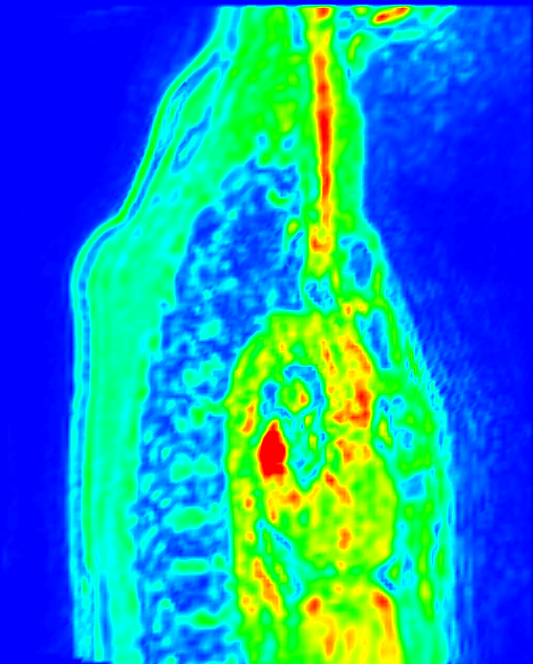}
\end{subfigure}
\begin{subfigure}{0.15\textwidth}
  \includegraphics[width=1\linewidth]{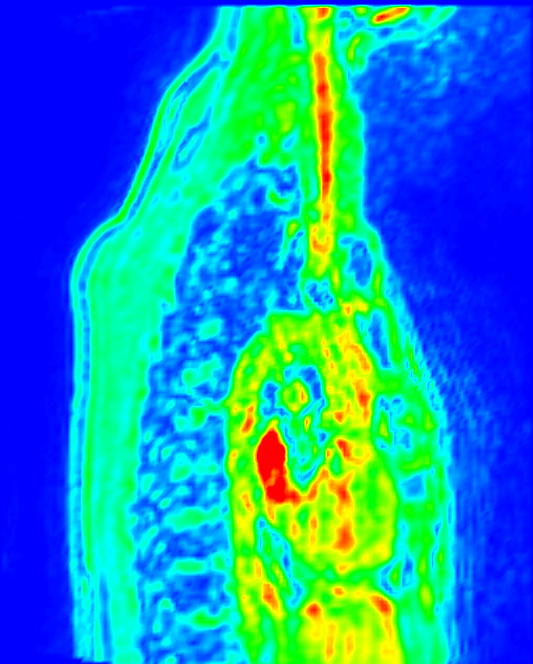}
\end{subfigure}
\caption{Every $4^{th}$ frame (from a total of 24 frames) of the magnitude component of a 4D Flow image.}
\label{fig:mag_frames}
\end{figure}

\begin{figure}[!ht]
\centering
\begin{subfigure}{0.15\textwidth}
  \includegraphics[width=1\linewidth]{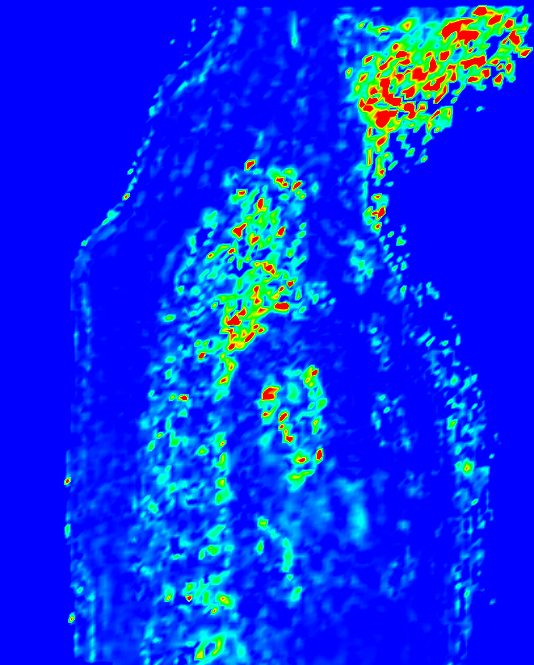}
\end{subfigure}
\begin{subfigure}{0.15\textwidth}
  \includegraphics[width=1\linewidth]{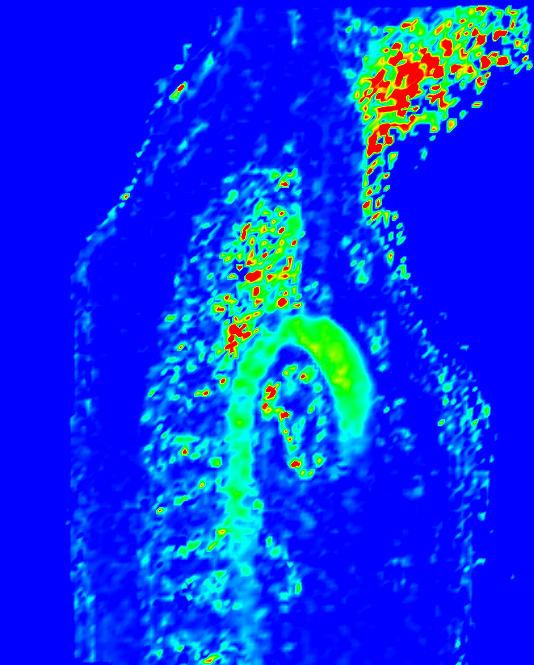}
\end{subfigure}
\begin{subfigure}{0.15\textwidth}
  \includegraphics[width=1\linewidth]{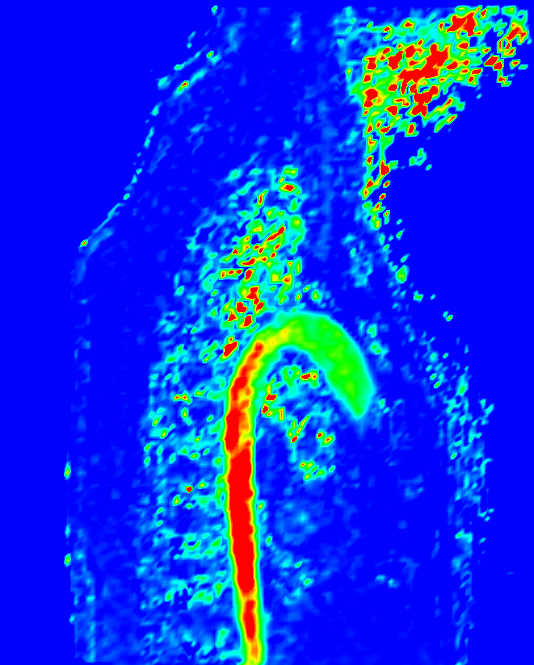}
\end{subfigure}
\begin{subfigure}{0.15\textwidth}
  \includegraphics[width=1\linewidth]{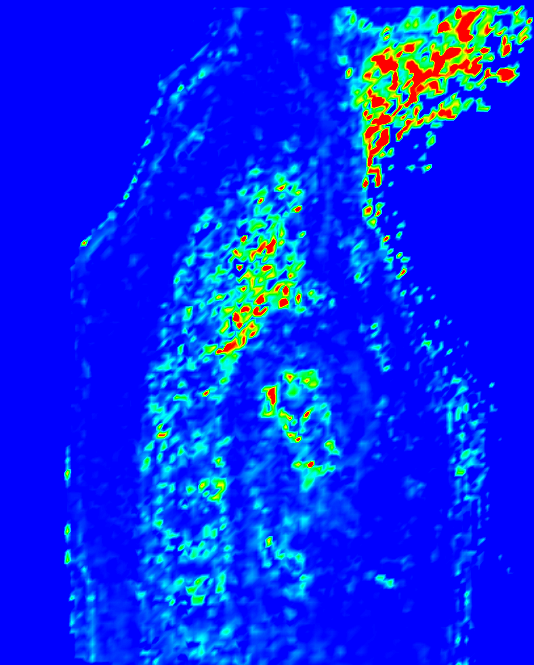}
\end{subfigure}
\begin{subfigure}{0.15\textwidth}
  \includegraphics[width=1\linewidth]{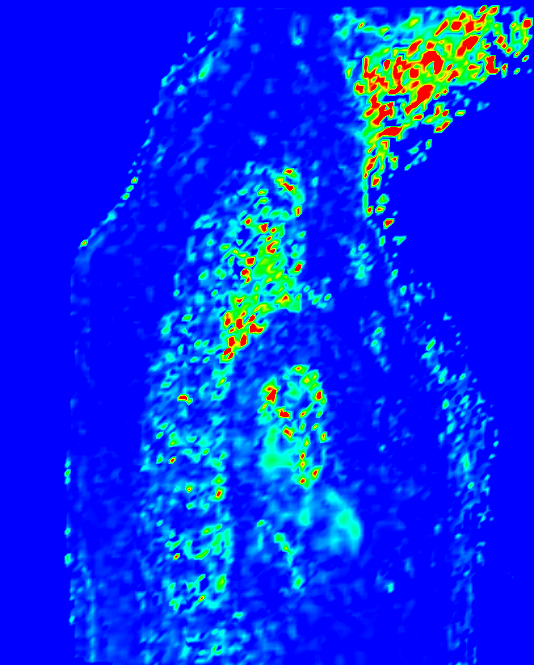}
\end{subfigure}
\begin{subfigure}{0.15\textwidth}
  \includegraphics[width=1\linewidth]{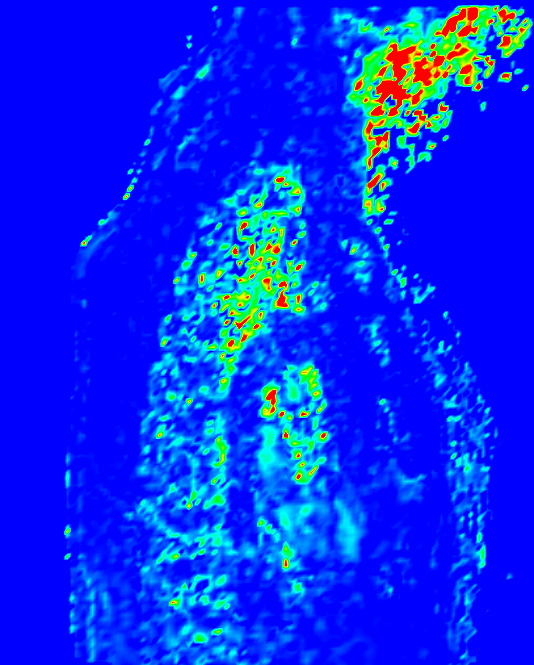}
\end{subfigure}
\caption{Every $4^{th}$ frame (from a total of 24 frames) of the speed (calculated from $\mathbf{v}$) of a 4D Flow image.}
\label{fig:speed_frames}
\end{figure}

A 4D Flow image is an example of a time-resolved image with codomain $\mathbb{R}^4$ (three spatial components and one components for the magnitude). Each frame of a 4D Flow image is a vector image with three components measuring fluid velocity and one component, ``magnitude", measuring water content. The frames of a 4D Flow image correspond to uniformly spaced times within a cardiac cycle (see Figure~\ref{fig:mag_frames} and Figure~\ref{fig:speed_frames}).

\begin{figure}[!ht]
\centering
\begin{subfigure}{0.15\textwidth}
  \includegraphics[width=1\linewidth]{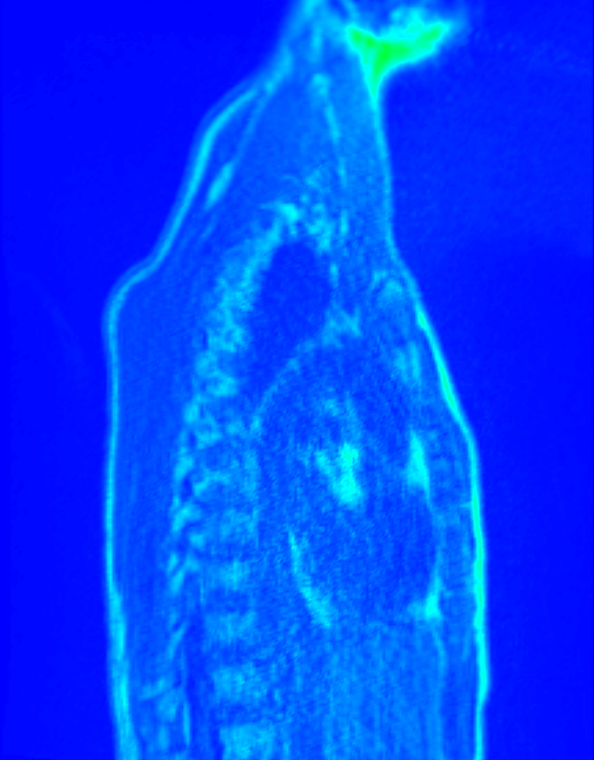}
\end{subfigure}
\begin{subfigure}{0.15\textwidth}
  \includegraphics[width=1\linewidth]{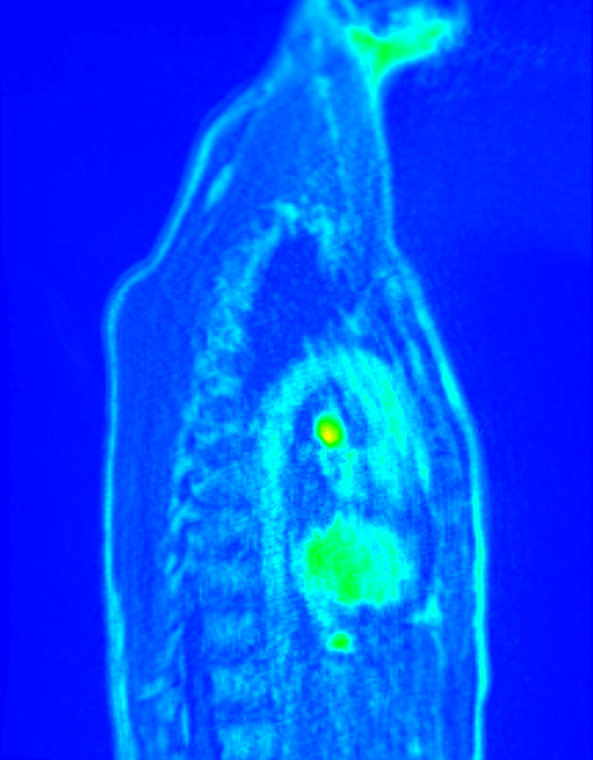}
\end{subfigure}
\begin{subfigure}{0.15\textwidth}
  \includegraphics[width=1\linewidth]{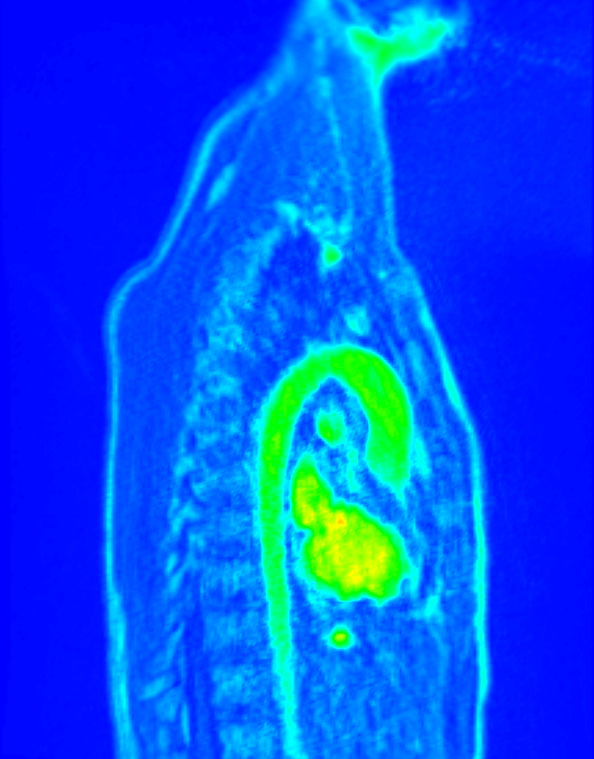}
\end{subfigure}
\begin{subfigure}{0.15\textwidth}
  \includegraphics[width=1\linewidth]{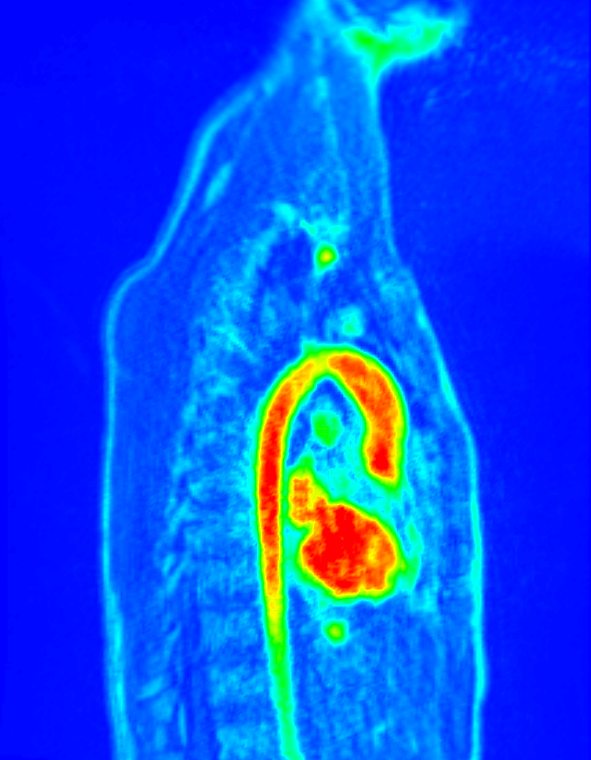}
\end{subfigure}
\begin{subfigure}{0.15\textwidth}
  \includegraphics[width=1\linewidth]{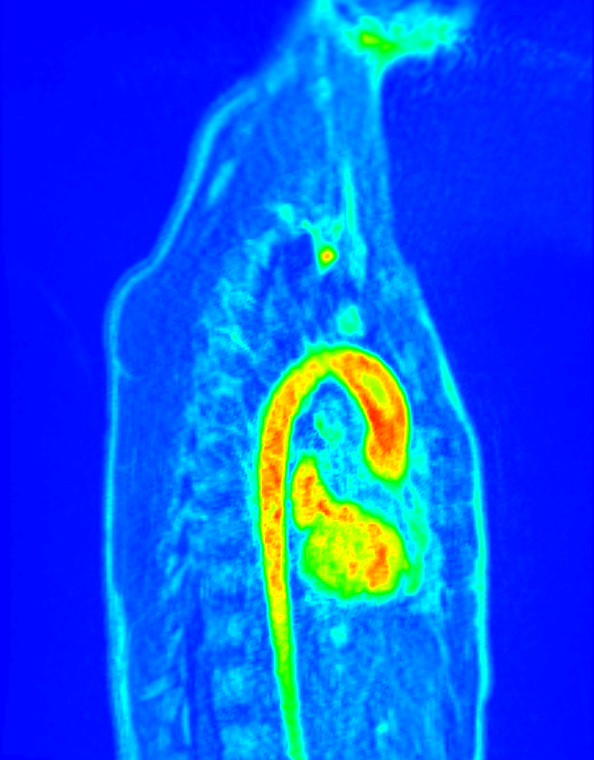}
\end{subfigure}
\begin{subfigure}{0.15\textwidth}
  \includegraphics[width=1\linewidth]{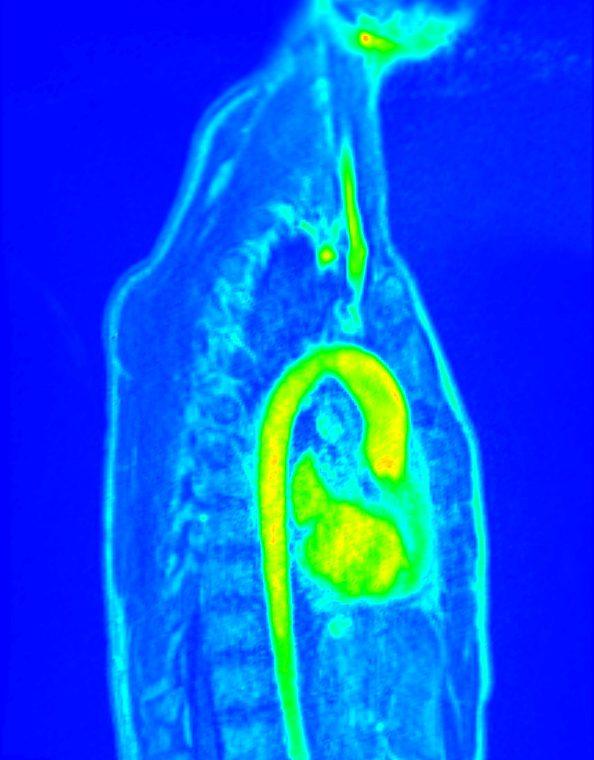}
\end{subfigure}
\caption{All 6 frames of an MRA scan at mid-diastole indicating progression of a contrast agent.}
\label{fig:mra_frames}
\end{figure}

Over the course of an MRA scan, a contrast agent may be injected into a subject before the beginning of the scan which perfuses through the circulatory system as the scan progresses. The duration of the scan covers many cardiac cycles, but each frame reflects the subject at \textit{a fixed phase} of a cardiac cycle. The scalar values $\rho$ of these frames are sensitive to water content and are amplified by the local concentration of contrast agent as the agent dissipates (see Figure~\ref{fig:mra_frames}).

\subsection{Segmentation}
The term \emph{segmentation} refers to both the product and the process of extracting an explicit surface representation from a scalar image. In this paper, segmentation is used to extract triangle meshes from MRA images representing tissue boundaries, in particular vessel network walls. 

\subsection{Scalarization of Vector Images} 
A vector image $\mathbf{v}$ describing the velocity of fluid traveling inside a vessel, implicitly encodes the inner surface of the vessel wall, but the usual segmentation software only extracts surfaces from scalar images. Two complementary methods to generate scalar images from vector images are the \emph{speed} scalarization and the \emph{coherence} scalarization images, described below. The desired mesh can then be segmented from either (or a combination of both) of these scalarized images.

\subsubsection*{The Speed Image of a Velocity Image}
The first method for scalarizing the velocity vector image is simply to replace each vector with its speed (i.e. Euclidean norm). The surface represented by a segmentation of the speed image represents the boundary between positive and near zero velocity flow\footnote{Due to the ``no-slip" condition, this surface is generally smaller than the actual vessel's inner surface.}.

\subsubsection{The Coherence Image of a Velocity Image} \label{sec:coherence}
A vector field $\mathbf{v} \colon \mathbb{R}^n \to \mathbb{R}^n$ is \emph{coherent} at a point $p$ if the directions of the velocities in a small neighborhood about $p$ are close to the direction of the velocity at $p$. More precisely:

\begin{equation}
    \text{coherence}(\mathbf{v},\mathbf{p}) := 1 - \lim_{\epsilon \to 0} \frac{\int_{B_\epsilon(\mathbf{p})} \cos{\theta_{\mathbf{p}}}}{|B_\epsilon|}
\end{equation}

Here, $B_\epsilon(\mathbf{p})$ denotes a ball of radius $\epsilon$ centered at $p$ and $\theta_{\mathbf{p}}$ is the function whose value at a point $\mathbf{q}$ is the angle between the velocities $\mathbf{v}(\mathbf{p})$ and $\mathbf{v}(\mathbf{q})$. \\

The velocity components of a single frame of a 4D Flow image (also denoted by $\mathbf{v}$) is a discrete form of a vector field. The corresponding discretized version of coherence is given by:

\begin{equation}
    \text{coherence}(\mathbf{v},\mathbf{p}) := 1 -  \frac{\sum_{\mathbf{q} \in N_1(\mathbf{p})} \cos{\theta_{\mathbf{p}}}(\mathbf{q})}{|N_1(\mathbf{p})|}
\end{equation}

where $N_1(\mathbf{p})$ is the set of voxels adjacent to $\mathbf{p}$.

The surface represented by a segmentation of the coherence image represents the boundary between coherent flow on the inside of the vessel and incoherent (i.e. noisy, turbulent, near stationary) flow on the outside of the vessel.

\begin{figure}[!ht]
\centering
\begin{subfigure}{0.3\textwidth}
    \scalebox{-1}[1]{
      \includegraphics[angle=90, width=1\linewidth]{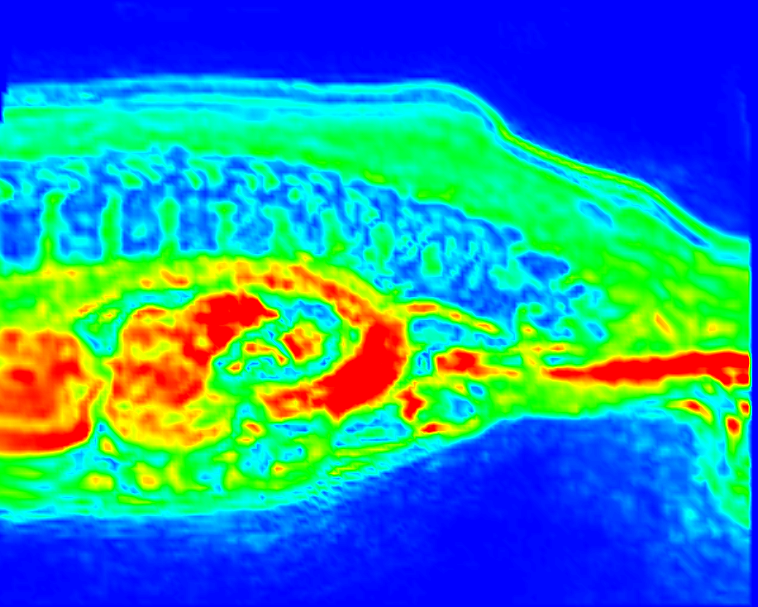}
    }       
\end{subfigure}
\begin{subfigure}{0.3\textwidth}
    \scalebox{-1}[1]{
      \includegraphics[angle=90, width=1\linewidth]{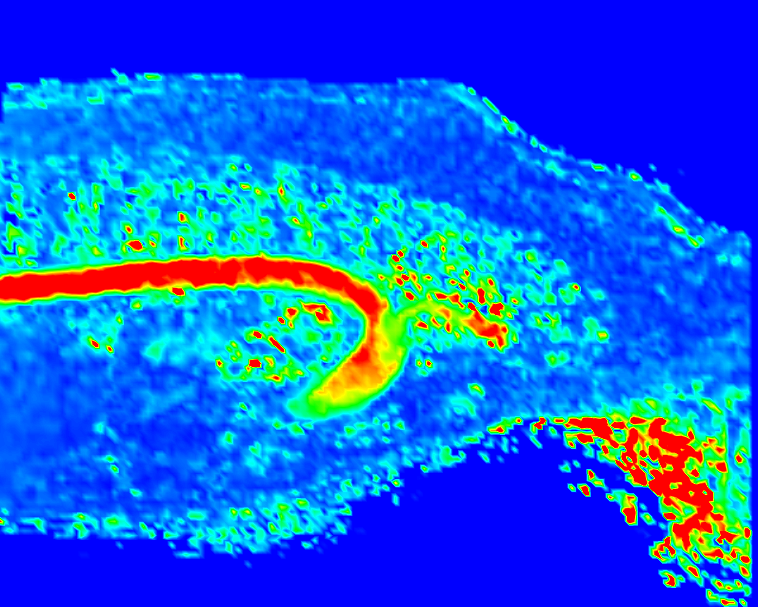}
    }       
\end{subfigure}
\begin{subfigure}{0.3\textwidth}
    \scalebox{-1}[1]{
      \includegraphics[angle=90, width=1\linewidth]{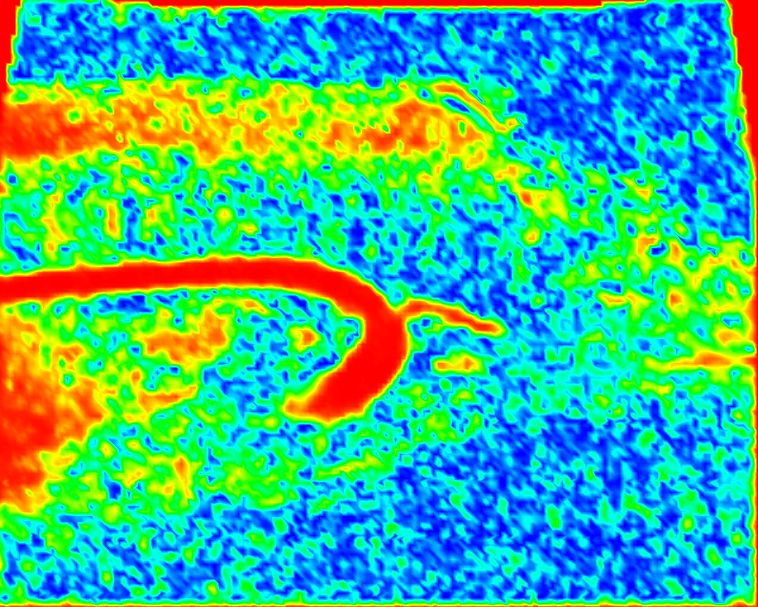}
    }       
\end{subfigure}
\caption{Scalar images extracted from a frame a 4D Flow image: (a) magnitude component, (b) speed of the velocity component and (c) coherence of the velocity component.}
\label{fig:scalarization}
\end{figure}

Figure~\ref{fig:scalarization} illustrates the speed and coherence scalarizations of the velocity component of a 4D Flow image. The corresponding magnitude component is also illustrated for comparison. For the particular illustrated example, it is apparent that the coherence captures the vessel wall most accurately than the other two scalar images. In fact, this was typical of all slices and 4D Flow images that were examined.

\subsection{Vessels, Centerlines, and their Networks}

Loosely speaking, a \emph{vessel} is a tubular surface. More precisely, the cross sections of the vessel perpendicular to its centerline are roughly circular and the radius of the cross sections are small compared to the length of the centerline. Under modest smoothness conditions a vessel is well-approximated by its centerline. Similarly, a smooth transformation of such a vessel is well-approximated by the transformation of its centerline. 

\emph{Centerline extraction} is the process whereby a polyline is constructed (usually from a segmentation of vessel) that approximates a centerline for the vessel. 
 
A \emph{vessel network} is simply a collection of vessels. The way that various vessels in the network intersect, branch, and join is irrelevant. A \emph{centerline network} is the corresponding collection of centerlines. 

\subsubsection{Registration of Polylines and Polyline Networks}

Ubiquitous in the field of image processing is the problem of \emph{registration}. Given two objects, $X_\text{fixed}$, $X_\text{floating}$, a transformation $T$ is sought so that $X_\text{registered} := T(X_\text{floating})$ aligns well with $X_\text{fixed}$. 

In the context of polylines, the objects of the registration are polylines and the transformations are polyline transformations. The task of registering of polyline networks is simply the construction of a single transformation that simultaneously registers all the individual polylines in the network. Details of this construction appear in Section ~\ref{sec:methods}. A polyline registration is called \emph{affine} (resp. \emph{rigid}) if the associated transformation is the restriction of an affine (resp. rigid) displacement map. 

A registration is deemed successful to the extent that $X_{registered}$ ``aligns better" with $X_\text{fixed}$ than does $X_\text{floating}$. The alignment of two polylines is quantified by their distance histograms (see Section \ref{sec:histogram}). In particular, one can consider the median value of the distance histograms; lower values of this median correspond to smaller ``distances" between centerlines and hence better alignments.  

\clearpage

\bibliography{main}
\bibliographystyle{spiejour}

\end{document}